\documentclass[%
 preprint,
 superscriptaddress,
 amsmath,amssymb,
 aps, physrev, english
]{revtex4-2}

\usepackage{graphicx}
\usepackage{dcolumn}
\usepackage{bm}

\usepackage[version=3]{mhchem}
\usepackage{acronym} 
\usepackage{longtable}
\usepackage{multirow}
\usepackage[T1]{fontenc}
\acrodef{AIMD}[AIMD]{ab initio molecular dynamics}
\acrodef{ARPES}[ARPES]{angle-resolved photoemission spectroscopy}
\acrodef{BZ}[BZ]{Brillouin zone}
\acrodef{DFT}[DFT]{density functional theory}
\acrodef{MCS}[MCS]{Monte Carlo sampling}
\acrodef{MLFF}[MLFF]{machine-learning force field}
\acrodef{MD}[MD]{molecular dynamics}
\acrodef{PAW}[PAW]{projector augmented-wave}
\acrodef{PBE}[PBE]{Perdew-Burke-Ernzerhof}
\acrodef{SOC}[SOC]{spin-orbit coupling}
\acrodef{SCF}[SCF]{self-consistent field}
\acrodef{ZG}[ZG]{Zacharias-Giustino}

\RequirePackage[normalem]{ulem}
\RequirePackage{color}
\definecolor{RED}{rgb}{1,0,0}
\definecolor{BLUE}{rgb}{0,0,1}
\definecolor{orange}{rgb}{1,0.65,0}
\definecolor{green}{rgb}{0,0.6,0}

\begin{document}


\title{Topological robustness of thermally disordered lattices: From average structures to ensemble electronic properties}

\author{Oleg Rubel}
\email{Contact author: rubelo@mcmaster.ca}
\affiliation{Department of Materials Science and Engineering, McMaster University, 1280 Main Street West, L8S 4L8, Hamilton, Ontario, Canada}

\author{Karekin Sadikian}
\affiliation{Department of Materials Science and Engineering, McMaster University, 1280 Main Street West, L8S 4L8, Hamilton, Ontario, Canada}

\date{\today}

\begin{abstract}
Band topology is usually assigned to a single crystal structure, yet at finite temperature a crystal is an ensemble of thermally disordered configurations with a fluctuating band gap.
We ask whether the thermally averaged structure is a faithful proxy for this ensemble, taking \ce{Bi2Se3} as a representative topological insulator and classifying band ordering in each \textit{ab initio} molecular dynamics snapshot by scaling the spin-orbit coupling.
The averaged structure fails in two ways: it misses the Rashba-like spin splitting that instantaneous symmetry breaking produces in individual configurations, and it substantially underestimates the band-gap renormalization, whereas a small ensemble of harmonic Monte Carlo configurations reproduces both the mean and the spread of the gap.
Using a band-inversion survival probability, we find that the band inversion of \ce{Bi2Se3} holds in every sampled configuration up to 600~K, above which individual configurations cross into normal band ordering well before the average gap closes.
Topological robustness at finite temperature is thus a property of the distribution over configurations rather than of any single structure.
\end{abstract}

\maketitle


\section{Introduction}

Finite temperature replaces a static crystal structure by an ensemble of thermally disordered configurations $\{\bm{R}\}$, where $\bm{R}$ denotes collective atomic coordinates.
The thermally averaged configuration $\langle \bm{R} \rangle_T$ does not necessarily yield the same electronic structure property $P$ as that obtained by averaging over instantaneous structures (different snapshots) \cite{Allen_JPCSSP_9_1976},
\begin{equation}\label{Eq: P(R)}
    P(\langle \bm{R} \rangle_T)
    \stackrel{?}{=}
    \langle P(\bm{R}) \rangle_T ,
\end{equation}
a distinction that forms the basis of the modern ensemble-averaged formulation of electron-phonon coupling \cite{Giustino_RMP_89_2017}.
Reported atomic coordinates from conventional experimental crystal structures determined by Bragg X-ray or neutron diffraction can reasonably be regarded as predominantly of the $\langle \bm{R} \rangle_T$ kind \cite{Allen_JPCSSP_9_1976}.
Crystal structures conventionally used as input for first-principles calculations, whether obtained from experiment or static structural optimization, predominantly represent an average or equilibrium crystal structure rather than instantaneous thermally disordered configurations.

Topological invariants in $\mathbb{Z}_2$ topological insulators are protected by time-reversal symmetry and depend on  crystal structure in a broad sense.
Thermal disorder can lead to breaking of spatial crystal symmetries, notably inversion, in instantaneous structures.
This raises a fundamental question: whether a band-topological classification based on the thermally averaged structure $\langle\mathbf R\rangle_T$ is applicable to the finite-temperature ensemble when the bulk gap fluctuates and can intermittently approach closure?
The finite-temperature robustness of the bulk topology underpins the existence of useful boundary states and limits the practical operating temperature of devices that rely on those states.

\citet{Garate_PRL_110_2013} provided a proof of principle for the impact of electron-phonon interactions on the band topology of insulators and semimetals from model Hamiltonians, with an emphasis on possibility of thermally driven topological crossovers.
\citet{Kim_PRB_92_2015} investigated the development of topological insulating phases in IV-VI compounds under dynamic lattice deformations using first-principles calculations and showed that nontrivial topological phases can be induced by selectively excited phonon modes.
\citet{Antonius_PRL_117_2016} and \citet{Monserrat_PRL_117_2016} used the first-principles electron--phonon coupling in real materials showing that the equilibrium thermal population of phonon modes can either promote \cite{Antonius_PRL_117_2016} or suppress \cite{Antonius_PRL_117_2016, Monserrat_PRL_117_2016} topologically non-trivial states depending on the material.
While \citet{Antonius_PRL_117_2016} neglected the thermal expansion, \citet{Monserrat_PRL_117_2016} showed that the thermal expansion makes an important contribution to the band gap renormalization in V-VI compound semiconductors, including \ce{Bi2Se3} topological insulator.
In \ce{Bi2Se3}, the $\Gamma$-point gap did not close within the calculated temperature range (0$-$600~K), which was interpreted as persistence of the inverted band ordering \cite{Monserrat_PRL_117_2016}.
Both works \cite{Antonius_PRL_117_2016, Monserrat_PRL_117_2016} relied on harmonic phonons.
Recently, \citet{Focassio_PRB_104_2021} used \ac{AIMD} and the real-space spin Bott index to study disorder-driven topological-to-trivial transition in \ce{Bi2Se3}.
The topological-to-trivial transition between 1500 and 1600~K was accompanied by the band gap closure and the crystalline-to-amorphous phase transition.
The \ac{AIMD} in Ref.~\cite{Focassio_PRB_104_2021} naturally captures anharmonic effects.
However, their evaluation of the Bott index required mapping the electronic structure onto a localized orbital basis.
Recent \ac{ARPES} measurements on amorphous \ce{Bi2Se3} have reported spin-momentum-locked surface states \cite{Corbae_NM_22_2023}, which can be interpreted as evidence that nontrivial electronic topology does not necessarily require long-range crystalline order.

Here we are going to study effect of the finite temperature on the electronic structure and the band inversion in \ce{Bi2Se3} as the model material using computationally efficient designed harmonic representations of thermal disorder.
We compare our results with \ac{AIMD}.
A comparison between Monte Carlo sampling of the harmonic vibrational distribution and \ac{AIMD} was made by \citet{Zacharias_PRB_102_2020} for the band-gap renormalization of Si and \ce{SrTiO3}, where the two nearly coincided for Si, whereas \ac{AIMD} gave a substantially larger renormalization for the strongly anharmonic \ce{SrTiO3}.
Whether a harmonic description is adequate depends on the vibrational amplitude and thus on temperature.
\ce{Bi2Se3} remains largely harmonic up to room temperature \cite{Kim_APL_100_2012, Tian_PRB_95_2017, Park_PRB_88_2013}, whereas the present study extends to 1050~K, so it is not obvious in advance whether a harmonic representation of thermal disorder can reproduce the \ac{AIMD} ensemble.
Here we extend the comparison to band ordering and its fluctuations, with the goal of identifying which designed thermally disordered structures come closest to \ac{AIMD} in reproducing both the thermally averaged electronic structure and the topology-sensitive fluctuations.

\section{Methods}\label{sec: Methods}

\subsection{Designed harmonic representations of thermal disorder}

We performed \ac{DFT} calculations (VASP version 6.5.1 \cite{Kresse_PRB_54_1996}) with \ac{PAW} potentials \cite{Kresse_PRB_59_1999} for Bi and Se with 5 and 6 valence electrons, respectively.
We created a 135-atom \ce{Bi2Se3} supercell of the size $3a \times 3a \times c$, where $a$ and $c$ are lattice parameters in the hexagonal setting.
Calculations were performed at the \ac{PBE} \cite{Perdew_PRL_77_1996} level of theory with the \citet{Grimme_JCP_132_2010} dispersion correction (D3 with zero damping).
The precision tag was set to `Accurate'.
The cutoff energy for the plane wave expansion was set at 212~eV, which is the largest of the maximum cutoffs of the \ac{PAW} potentials.
\Ac{SOC} was excluded in structure relaxation and phonon calculations.
The finite-temperature lattice parameters were taken from the \ac{AIMD} results (see below) and fixed.
Internal degrees of freedom were relaxed with the convergence energy parameter for structural changes of $10^{-6}$~eV (the \ac{SCF} energy convergence was $10^{-8}$~eV) that resulted in forces of less than $10^{-4}$~eV/{\AA}.
Smearing for band occupancies was Gaussian with the width of 0.01~eV.
The \ac{BZ} was sampled using a $\Gamma$-centred $k$-mesh with spacing of 0.15~{\AA}$^{-1}$, which produces the mesh of $4 \times 4 \times 2$ in the supercell \ac{BZ}.

The relaxed structure with lattice parameters from $NpT$ \ac{AIMD} was used to generate thermally disordered structures (either \ac{ZG} \cite{Zacharias_PRB_94_2016} or \ac{MCS} \cite{Karsai_NJP_20_2018} structures) based on harmonic phonon modes calculated using a finite displacement method implemented in VASP.
The phonon calculation used the step size of 0.015~{\AA} and the \ac{SCF} energy convergence of $10^{-6}$~eV.
Such displacements introduced total energy change of the order of $5\times 10^{-4}$~eV, which is significantly greater than the convergence energy parameter for structural relaxation of $10^{-6}$~eV (not to be confused with the \ac{SCF} energy convergence).
The VASP implementation uses the same $T$- and frequency-dependent quantum-harmonic Gaussian displacement width $\sigma_{\nu}(T)$ of the normal-mode coordinate for both \ac{ZG} and \ac{MCS} structures with only one distinction.
It is related to how displacements of phonon modes are scaled when constructing the thermally disordered structures.
The \ac{ZG} algorithm assigns alternating signs ($+, -, +, -, \ldots$) to the normal-mode displacements in ascending order of eigenfrequencies.
While \ac{MCS} uses a standard normal distribution as the scaling factor for $\sigma_{\nu}(T)$ drawn independently for each normal mode and each configuration.
As a result, \ac{ZG} generates only one disordered structure for each $T$, while \ac{MCS} produces multiple structures (we limited our study to 10 random structures).
It should be noted that the VASP \ac{ZG} implementation applies the alternating signs to the eigenvectors as returned by the diagonaliser, without the gauge fixing prescribed in Ref.~\citenum{Zacharias_PRB_94_2016}.

\subsection{\textit{Ab initio} molecular dynamics}

\Ac{AIMD} was performed in two steps.
We first performed isothermal-isobaric ($NpT$) ensemble simulations to determine the lattice parameters at a given temperature, followed by canonical $NVT$ ensemble simulations at fixed lattice parameters to extract snapshots for subsequent electronic-structure analysis.

For the $NpT$ simulations, the same computational settings as for the structural calculations were used, including the \ac{PAW} potentials, plane-wave cutoff energy, precision setting, exchange-correlation functional, Gaussian smearing, and $k$-point sampling.
Crystal symmetry was disabled to allow unrestricted thermal displacements.
Constant-temperature $NpT$ \ac{MD} was performed at zero applied external pressure using a Langevin thermostat and barostat \cite{Parrinello_PRL_45_1980, Parrinello_JAP_52_1981}.
Each simulation consisted of 20,000 steps with a time step of 2~fs, corresponding to a total simulation time of 40~ps.
The electronic energy convergence criterion was set to $10^{-6}$~eV.
The atomic and lattice degrees of freedom were coupled to the Langevin thermostat using friction coefficients of 10~ps$^{-1}$.
The fictitious mass associated with the lattice degrees of freedom was set to 1000~amu.
The relatively large lattice mass moderates the response of the simulation cell to instantaneous stress fluctuations, while the lattice friction coefficient controls damping of the cell dynamics.
The stress tensor was evaluated during \ac{MD} and used to drive fluctuations of the simulation cell.

To accelerate the \ac{MD} simulations, the \ac{MLFF} implemented in VASP was trained on the fly \cite{Jinnouchi_PRB_100_2019}.
An independent \ac{MLFF} was trained from scratch for each simulation using the element-reduced descriptor.
The force-error threshold for selecting new first-principles training configurations was updated dynamically based on the history of Bayesian force uncertainties.
The updated threshold was set to 90\% of the historical average Bayesian force uncertainty, thereby promoting more frequent acquisition of first-principles reference data.
Local atomic environments with an estimated force error exceeding 40\% of the current force-error threshold were considered for inclusion in the sparse kernel representation.
These relatively conservative settings favor more frequent first-principles sampling and a more inclusive selection of local reference configurations during \ac{MLFF} training.
Root-mean-square errors relative to the first-principles training data increased with temperature from 0.02 to 0.08~eV/{\AA} for forces and from 0.2 to 0.6~kbar for stresses, and the \ac{MLFF} replaced explicit electronic-structure calculations at 97$-$99\% of the steps.

The $NVT$ simulations used the same electronic-structure and \ac{MLFF} settings as the $NpT$ simulations.
The simulation cell was kept fixed, while crystal symmetry remained disabled to allow unrestricted thermal displacements of the atoms.
Temperature was controlled using the canonical sampling through velocity rescaling thermostat \cite{Bussi_JCP_126_2007}.
The characteristic time scale of the thermostat was set to 80 \ac{MD} steps, corresponding to 160~fs for the time step of 2~fs.
This relatively long coupling time provides gradual temperature control while allowing the intrinsic atomic dynamics to evolve between substantial thermostat corrections.
Each $NVT$ simulation consisted of 20,000 \ac{MD} steps, corresponding to a total simulation time of 40~ps.

\subsection{Spin-orbit coupling scaling}

The original VASP code was modified to introduce a scaling parameter $\lambda$ for \ac{SOC}, where $\lambda=0$ corresponded to the absence of \ac{SOC} and $\lambda=1$ to its full physical strength.
By gradually varying $\lambda$ in a series of \ac{SCF} calculations, we tracked the closure and subsequent reopening of the band gap associated with \ac{SOC}-driven band inversion, following the spirit of the earlier Ref.~\citenum{Wang_PRB_83_2011}.
\citet{Wang_PRB_83_2011} demonstrated switching of the orbital character while passing a critical value of $\lambda$ associated with band inversion in \ce{Bi2Te2S}, and examined the resulting surface states in a slab model.
In addition, \citet{GomezBastidas_CPC_292_2023} considered \ce{Bi2Se3} at full ($\lambda=1$) and substantially reduced ($\lambda\simeq0.21$) \ac{SOC} strengths and established the corresponding non-trivial and trivial phases, respectively, through hybrid Wannier charge center analysis.

\section{Results and discussion}

\subsection{Thermal expansion}

The fully optimized structure of \ce{Bi2Se3} was used as the initial configuration for the $NpT$ simulations.
During the $NpT$ dynamics, the lattice parameters were allowed to evolve.
Their convergence was monitored using the forward cumulative average
\begin{equation}
    \langle x \rangle_n = \frac{1}{n}\sum_{i=1}^{n} x_i,
    \qquad n \leq N ,
\end{equation}
where $x_i$ is the value of a parameter at the $i$th \ac{MD} step, $n$ is the number of steps included in the cumulative average, and $N$ is the total number of steps in the \ac{MD} trajectory.

\begin{figure}[h]
    \centering\includegraphics[width=0.5\textwidth]{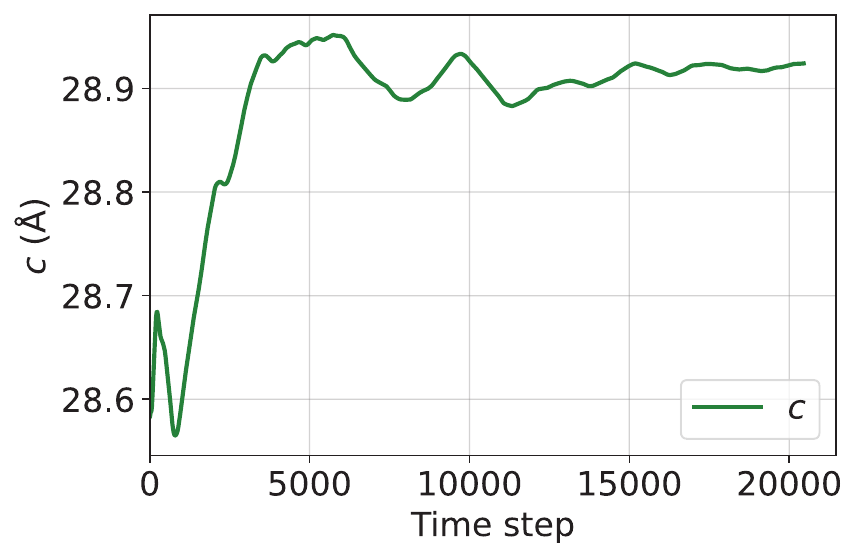}
    \caption{Evolution of the forward cumulative average lattice parameter $c$ during the $NpT$ \ac{AIMD} simulation at the temperature of 600 K.}
    \label{fig-lattice-param-NPT-600K-cumulative-time-average}
\end{figure}

Figure~\ref{fig-lattice-param-NPT-600K-cumulative-time-average} shows the evolution of the forward cumulative average of the out-of-plane hexagonal lattice parameter $c$ at 600~K.
The cumulative average becomes stable after approximately 15,000 time steps, corresponding to a simulation time of 30~ps for a time step of 2~fs.
The finite-temperature lattice parameter was determined by averaging over the entire \ac{MD} trajectory, including the initial equilibration period.

Once the finite-temperature lattice parameters were determined, the subsequent $NVT$ simulations were performed using a fixed simulation cell with the hexagonal symmetry restored, such that $a=b$, $c\ne a$, $\alpha=\beta=90^\circ$, and $\gamma=120^\circ$.
As an additional check of the resulting lattice parameters, the stress tensor was monitored during the $NVT$ simulations.
The magnitude of each diagonal component of the stress tensor, averaged over the \ac{MD} trajectory, was below 1~kbar.

\begin{figure}[h]
    \centering\includegraphics[width=\textwidth]{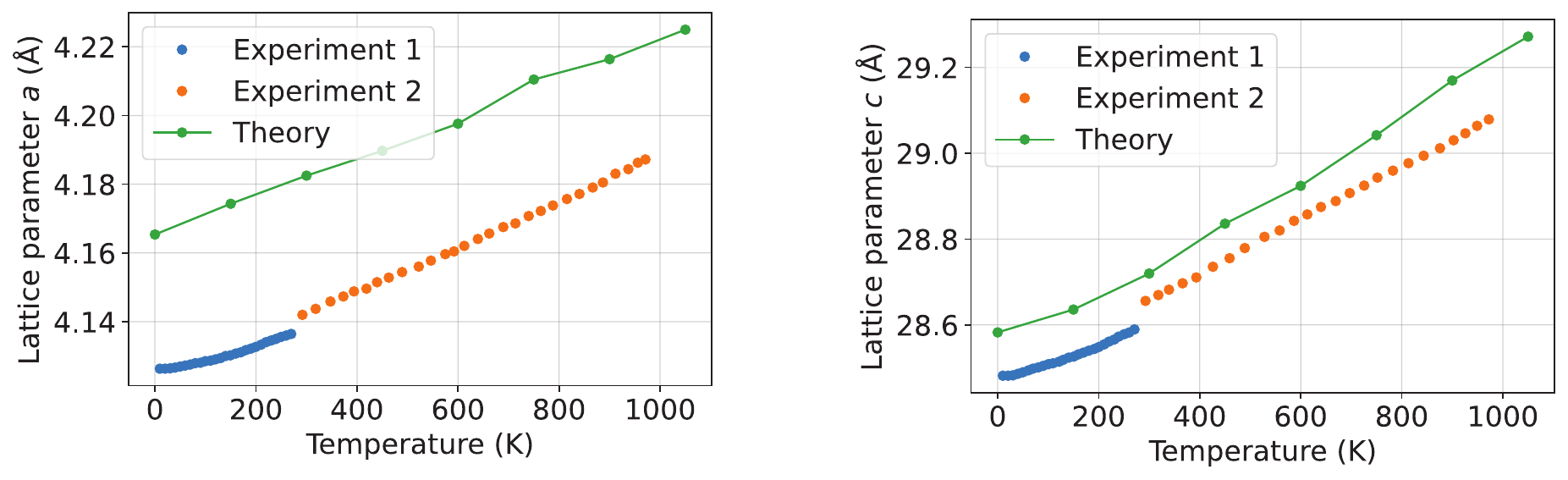}
    \caption{Temperature-dependent lattice parameters evaluated from \ac{AIMD} and compared with experiment.
    Experimental data were digitized from Refs.~\citenum{Chen_APL_99_2011, Kevy_JSSC_330_2024} (labelled as 1 and 2, respectively).}
    \label{fig-lattice-param-vs-T}
\end{figure}

Figure~\ref{fig-lattice-param-vs-T} compares the temperature dependence of the lattice parameters obtained from \ac{AIMD} with experimental values.
The offset between theory and experiment is of the order of 1\%, while the slopes $da/dT$ and $dc/dT$ are overall in good agreement with experiment.
This agreement serves as a quantitative validation of the structural model in terms of the external degrees of freedom and their temperature evolution.
The calculated lattice parameters at 0~K remain within 1.5\% of experiment after increasing the plane-wave cutoff energy by 25\% and including \ac{SOC} in the structural relaxation.
The self-consistent structure relaxed with the expanded basis set and \ac{SOC} has the band gap of 0.32~eV at $\Gamma$, which is 0.04~eV greater than our production calculations quoted below in Sec.~\ref{Sec:Results:Topological robustness}.
The increased gap at 0~K, i.e., the stronger band inversion obtained with these settings, suggests that the production calculations may provide a more conservative assessment of the thermal stability of the inverted band structure.

It is worth noting that \citet{Monserrat_PRL_117_2016} reported that plain \ac{PBE}  (no dispersion correction is reported in their methods) significantly overestimated both the equilibrium volume and thermal expansion of \ce{Bi2Se3}.
Consequently, they used experimental lattice parameters up to approximately 250~K and extrapolated them to higher temperatures.
In the present work, the finite-temperature lattice parameters are instead obtained directly from $NpT$ \ac{AIMD}.
Our values agree much better with experiment over the experimentally accessible temperature range.
The highest temperature considered, 1050~K, lies above the melting point of \ce{Bi2Se3} (983~K \cite{Kevy_JSSC_330_2024}), and the corresponding results describe a superheated crystal that remains ordered on the time scale of the simulation.

\subsection{Finite-temperature band structure}

\begin{figure}[h]
    \centering\includegraphics[width=0.35\textwidth]{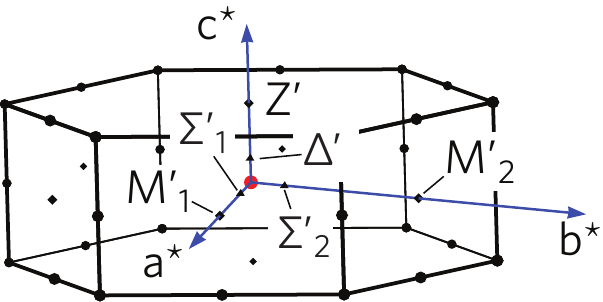}
    \caption{Brillouin zone of the $3 \times 3 \times 1$ supercell.
    The auxiliary points $\Sigma_1'$ and $\Sigma_2'$, located at 12\% of the corresponding $\Gamma-M_1'$ and $\Gamma-M_2'$ distances, and $\Delta'$, located at 32\% of the $\Gamma-Z'$ distance, define the reciprocal-space paths used for detailed sampling of the band structure in the vicinity of $\Gamma$.}
    \label{fig-BZ-supercell}
\end{figure}

We begin the finite-temperature band-structure analysis by identifying the high-symmetry points in the \ac{BZ} of the supercell.
Since the band inversion in \ce{Bi2Se3} occurs at the $\Gamma$ point, three auxiliary $k$ points were introduced in its vicinity along the reciprocal-lattice-vector directions to define shorter reciprocal-space paths for detailed sampling of the electronic structure near $\Gamma$.
This additional sampling is important for thermally distorted structures.
Breaking the inversion symmetry allows spin splitting away from $\Gamma$, and the minimum band gap can consequently shift away from the $\Gamma$ point, with larger deviations possible for structures exhibiting stronger thermal distortions.
Additional motivation to sample $k$ points in the vicinity of $\Gamma$ comes from the fact that the top of the valence band in \ac{PBE} is located approximately midway along the $\Gamma\!-\!M'_1$ path, which is an artifact (not shown here).
According to experiment and quasiparticle calculations \cite{Nechaev_PRB_87_2013}, the top of the valence band is at $\Gamma$.
Figure~\ref{fig-BZ-supercell} shows the \ac{BZ} of the supercell, including the high-symmetry points and the three auxiliary points $\Sigma_1'$, $\Sigma_2'$, and $\Delta'$ located near $\Gamma$.
Throughout this work, primed labels are used for $k$ points in the supercell \ac{BZ} to distinguish them from the corresponding notation for the primitive-cell \ac{BZ}.

We first define how the finite-temperature electronic structure is sampled.
The most direct representation of the thermal ensemble is obtained from the $NVT$ \ac{AIMD} trajectories.
Each 20,000-step trajectory was divided into 42 approximately equal intervals of 476--477 \ac{AIMD} steps, corresponding to approximately 0.95~ps per interval.
Every second interval was retained for sampling ($2,4,\ldots,42$), and one frame was selected randomly from each retained interval.
This procedure yielded 21 configurations per temperature with a guaranteed separation of at least 0.95~ps, thereby reducing temporal correlation between the selected configurations.
In addition to the \ac{AIMD} configurations, we considered 10 structures generated using \ac{MCS} of the quantum-harmonic vibrational distribution and a single representative structure generated using the \ac{ZG} approach.
Finally, a time-averaged structure was constructed by averaging the atomic coordinates over the \ac{MD} trajectory.
This structure provides an approximation to the average structural information that would typically be obtained from diffraction experiments.

\begin{figure}[h]
    \centering\includegraphics[width=\textwidth]{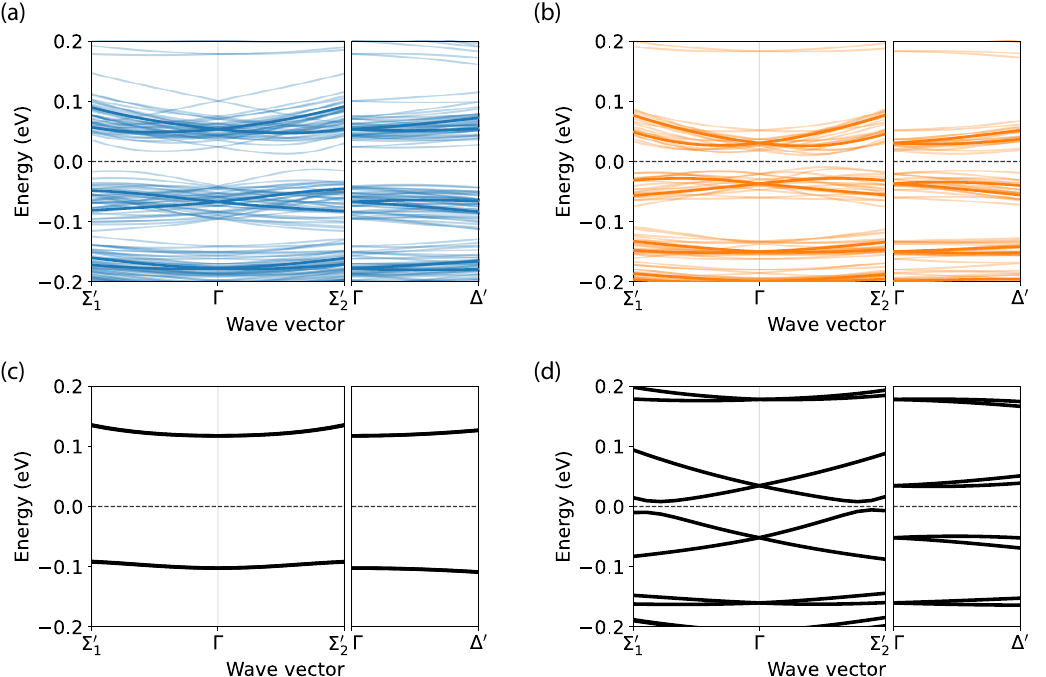}
    \caption{Band structure in the vicinity of $\Gamma$ at 600~K obtained using different structural sampling methods: (a) 21 individual \ac{AIMD} snapshots, (b) 10 individual \ac{MCS} configurations, (c) the thermally averaged structure, and (d) a single representative \ac{ZG} structure.
    For \ac{AIMD} and \ac{MCS}, individual configurations are shown by semi-transparent lines and the thick lines represent the configuration-averaged band energies.
    \Ac{SOC} is included.}
    \label{fig-band-struct-600K-sampling-methods}
\end{figure}

Figure~\ref{fig-band-struct-600K-sampling-methods} shows the band structures at 600~K obtained using the different structural sampling methods.
\Ac{AIMD} and \ac{MCS} are represented by multiple configurations, and the band structure of each structural realization is shown by thin semi-transparent lines in panels (a) and (b).
The thick lines show the configuration-averaged band energies, obtained by averaging eigenvalues with the same band index at each $k$ point.
Among the reduced harmonic representations, \ac{MCS} shows the closest agreement with the \ac{AIMD} ensemble.
The one-shot \ac{ZG} structure produces a stronger reduction of the band gap and places the system closer to gap closure.
In contrast, the band structure obtained from the thermally averaged crystal structure $\langle \bm{R} \rangle_{\textrm{600~K}}$ shows the least resemblance to the \ac{AIMD} ensemble, pointing towards inequality in Eq.~\eqref{Eq: P(R)}.

For band structures shown in Fig.~\ref{fig-band-struct-600K-sampling-methods}(a), (b), and (d), the band gap at $\Gamma$ is greater than the minimum band gap, which is displaced away from $\Gamma$.
\citet{Monserrat_PRL_117_2016} used non-closure of the temperature-renormalized $\Gamma$-point gap to infer the persistence of topological order in \ce{Bi2Se3}.
In the instantaneous thermally distorted structures, inversion-symmetry breaking leads to Rashba-like spin splitting away from $\Gamma$.
In contrast, the time-averaged structure is effectively centrosymmetric and does not exhibit spin splitting [Fig.~\ref{fig-band-struct-600K-sampling-methods}(c)].
A general mechanism for phonon-assisted Rashba-like spin splitting in centrosymmetric materials was proposed by \citet{Monserrat_arXiv_1711.06274} and demonstrated using first-principles calculations for \ce{CsPbCl3}.
Related effects associated with thermally induced local symmetry breaking have been studied in lead-halide perovskites \cite{McKechnie_PRB_98_2018, Niesner_PNAS_115_2018}.

\begin{figure}[h]
    \centering\includegraphics[width=\textwidth]{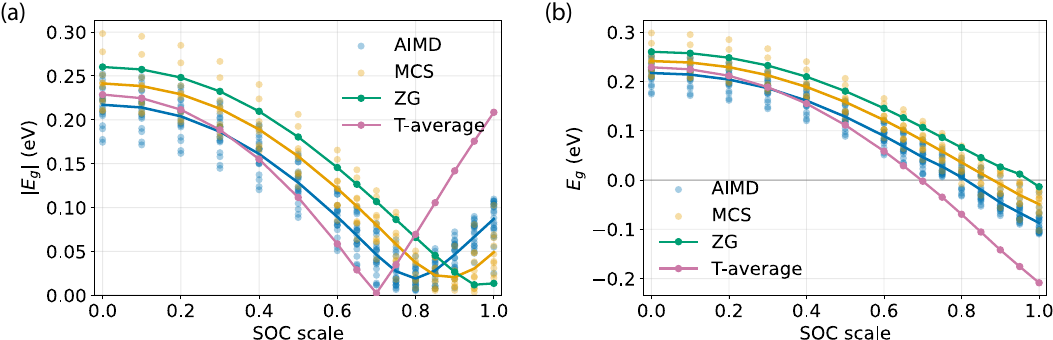}
    \caption{Absolute (a) and signed (b) band gaps at 600~K as a function of the \ac{SOC} scale.
    The minimum band gap was evaluated in the vicinity of $\Gamma$, specifically along $\Sigma'_1\!-\!\Gamma\!-\!\Sigma'_2$ and $\Gamma\!-\!\Delta'$ $k$-point paths in the supercell \ac{BZ} in Fig.~\ref{fig-BZ-supercell}.
    The negative gap indicates inverted band ordering.
    Various structural models were considered: individual \ac{AIMD} snapshots, \ac{MCS} configurations, the thermally averaged structure ($T$-average), and the single representative \ac{ZG} structure.
    When multiple structures are present, individual configurations are plotted with semi-transparent points overlaid with the configuration-averaged trend line.}
    \label{fig-Eg-vs-SOC-scale}
\end{figure}

We next determine whether a particular thermally sampled atomic configuration preserves the band inversion or evolves toward normal band ordering.
For this purpose, we use \ac{SOC} scaling.
Figure~\ref{fig-Eg-vs-SOC-scale}(a) shows the absolute band gaps at 600~K as a function of the \ac{SOC} scale $\lambda$ for the different structural sampling methods.
In general, the band gap decreases with increasing $\lambda$, approaches a minimum, and subsequently increases again.
For the time-averaged structure, for which the band gap occurs at $\Gamma$, an actual gap closure is observed.
For \ac{AIMD} and \ac{MCS}, individual configurations are shown by semi-transparent points and the configuration-averaged band gaps are shown by solid lines.
The decrease and subsequent reopening of the gap with increasing \ac{SOC} strength is taken as a signature of inverted band ordering.
The crossover between normal and inverted band ordering takes place at $\lambda = 0.6$ at 0~K (not shown here), shifting to $\lambda \approx 0.7\ldots0.95$ at 600~K depending on the thermal sampling method.
Figure~\ref{fig-Eg-vs-SOC-scale}(a) also illustrates fluctuations in both the band gap and the crossover value of $\lambda$ among the \ac{AIMD} snapshots, which are also captured by the \ac{MCS} structures.
These fluctuations, together with the shift of the crossover toward the full \ac{SOC} strength, indicate increasing proximity to a regime in which thermal fluctuations can drive individual configurations between normal and inverted band ordering.

To quantify the strength of band inversion, a signed band gap was reconstructed from the calculated absolute gap.
The signed gap was assumed to decrease monotonically with increasing \ac{SOC} strength.
The minimum in the absolute gap was identified from the onset of gap reopening, and the sign was reversed across this minimum.
Because the discrete sampling of $\lambda$ and $k$-space does not generally resolve the exact gap-closing point, the minimum-gap point was assigned to either the positive or negative branch according to which assignment minimized the local discontinuity in the slope of the reconstructed signed gap.
The resulting signed gaps are shown in Fig.~\ref{fig-Eg-vs-SOC-scale}(b), where negative values indicate inverted band ordering.
At full \ac{SOC} strength, the time-averaged structure exhibits the largest inverted gap, followed by \ac{AIMD} and \ac{MCS}.
The \ac{ZG} structure lies closest to gap closure and therefore exhibits the weakest band inversion among the structural models considered.
This result once again demonstrates that the equality in Eq.~\eqref{Eq: P(R)} does not hold, with \ac{MCS} providing the closest proxy for the \ac{AIMD} ensemble.

\subsection{Topological robustness at finite $T$}\label{Sec:Results:Topological robustness}

\begin{figure}[h]
    \centering\includegraphics[width=0.5\textwidth]{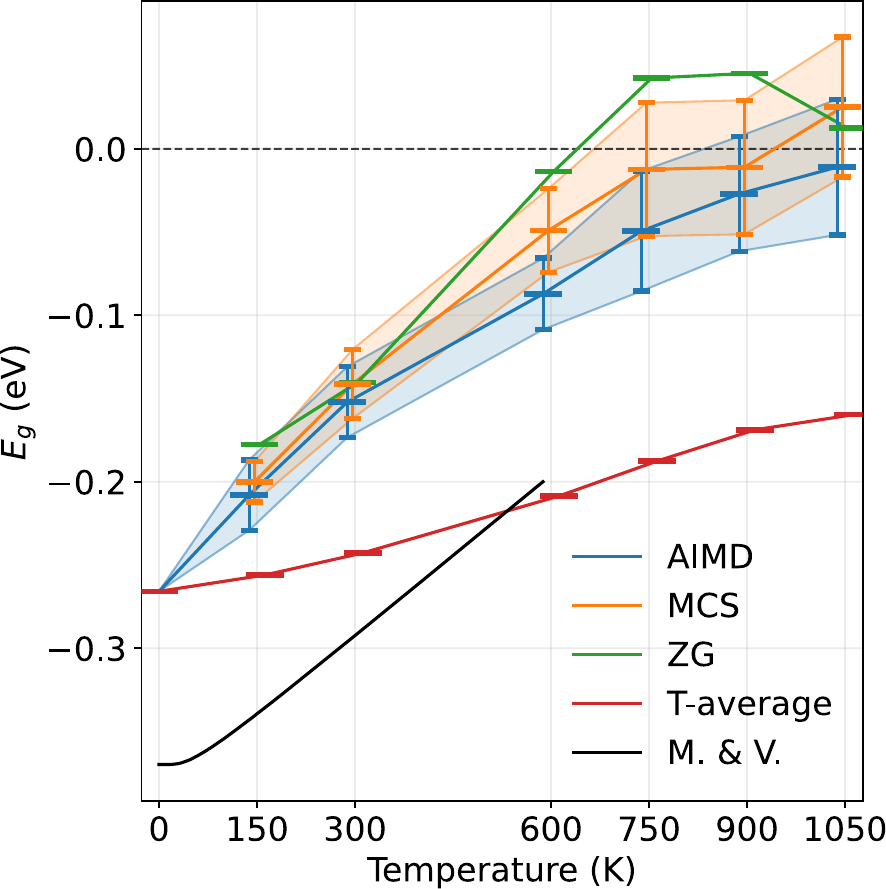}
    \caption{Signed band gaps at the physical \ac{SOC} scale as a function of temperature for various structural models.
    The minimum band gap was evaluated in the vicinity of $\Gamma$, specifically along $\Sigma'_1\!-\!\Gamma\!-\!\Sigma'_2$ and $\Gamma\!-\!\Delta'$ $k$-point paths in the supercell \ac{BZ} in Fig.~\ref{fig-BZ-supercell}.
    When multiple structures are present, the configuration-averaged values are supplemented with error bars and shaded bands representing $\pm 1$ standard deviation.
    $T$-average denotes the thermally averaged structure.
    Data digitized from \citet{Monserrat_PRL_117_2016} (M.~\&~V., thermal expansion combined with electron-phonon coupling) are included for comparison.
    Data points at finite temperatures are slightly offset horizontally for clarity.}
    \label{fig-Eg-vs-T}
\end{figure}

According to the \ac{AIMD} results presented in Fig.~\ref{fig-Eg-vs-T}, the initially inverted band gap progressively decreases in magnitude with increasing temperature and exhibits increasing thermal broadening.
At high temperatures, the distribution approaches the boundary between inverted and normal band ordering, suggesting the onset of a configuration-dependent crossover in the range of approximately 750--900~K.
The thermally averaged structures, which include thermal expansion but miss thermal disorder, capture only part of the weakening of the band inversion.
This is consistent with the earlier finding of \citet{Monserrat_PRL_117_2016} that thermal expansion and electron-phonon coupling make comparable contributions to the temperature dependence of the band gap.
Among the structures that incorporate thermal fluctuations of the atomic positions, \ac{MCS} configurations best reproduce the \ac{AIMD} results, including the thermal broadening of $E_g$, while predicting somewhat stronger weakening of the band inversion with temperature.
The single \ac{ZG} structures generally overestimate the thermal effect.
The accuracy of the \ac{ZG} mode-sign construction is expected to improve in the large-supercell limit \cite{Zacharias_PRB_94_2016}, while the $3\times3\times1$ supercell used here may retain finite-supercell effects.

Our 0~K gap at $\Gamma$ of 0.28~eV is within the spread of the experimental bulk band gap of \ce{Bi2Se3}, the magnitude of which depends on the measurement technique.
\Ac{ARPES} measurements estimate the band gap to be $\sim0.3$~eV \cite{Xia_NP_5_2009} and 0.33~eV \cite{Nechaev_PRB_87_2013}, while optical measurements give smaller values (in eV) of 0.18 \cite{Kohler_PSSB_63_1974}, 0.19 \cite{Post_PRB_88_2013}, 0.20 \cite{Orlita_PRL_114_2015}, 0.21 \cite{Gobrecht_ZP_190_1966}, and 0.22 \cite{Martinez_SR_7_2017}.
The \ac{DFT} band gaps reported by \citet{Monserrat_PRL_117_2016} are larger than our results (see Fig.~\ref{fig-Eg-vs-T}), likely due to differences in the protocol used to construct the structures and, in part, because we consider the minimum band gap in the vicinity of $\Gamma$, whereas \citet{Monserrat_PRL_117_2016} consider the gap at $\Gamma$ only.
Interestingly, after expressing their results using our signed-gap convention, the temperature dependence of their combined thermal-expansion and electron-phonon result is consistent with that obtained from \ac{AIMD}.
Broadening of the band-gap distribution was not reported in Ref.~\citenum{Monserrat_PRL_117_2016}.

\begin{figure}[h]
    \centering\includegraphics[width=0.5\textwidth]{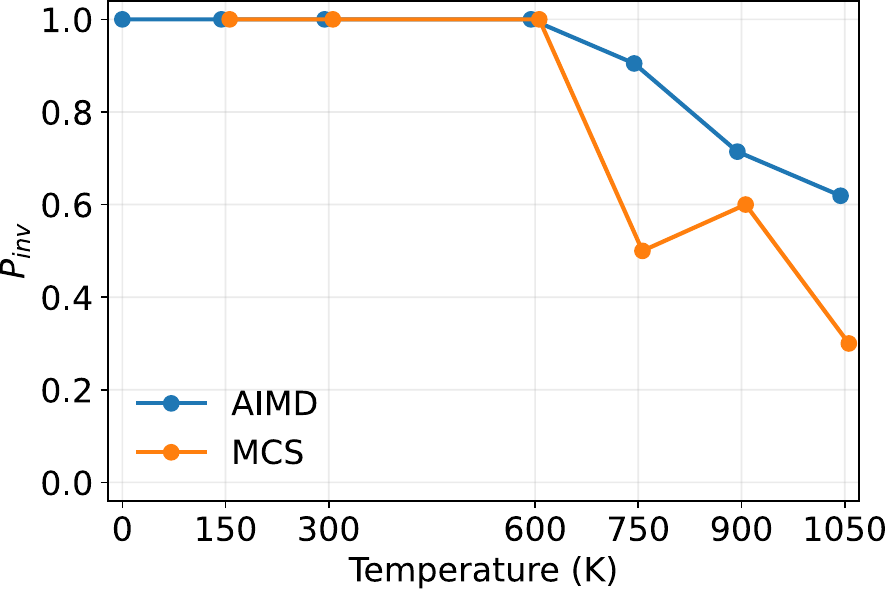}
    \caption{Band-inversion survival probability $P_{\mathrm{inv}}$ [Eq.~\eqref{Eq:Pinv}], i.e., the fraction of structural configurations with negative signed minimum band gaps near $\Gamma$, corresponding to inverted band ordering, as a function of temperature for \ac{AIMD} and \ac{MCS}.}
    \label{fig-band-inversion-prob}
\end{figure}

The band-gap magnitude and its thermal broadening in Fig.~\ref{fig-Eg-vs-T} already provide an indication of the stability of the band inversion as the distribution approaches zero.
To make the statistics more transparent, we define the band-inversion survival probability as
\begin{equation}\label{Eq:Pinv}
    P_{\mathrm{inv}}(T)=\Pr(E_g<0),
\end{equation}
corresponding to the fraction of sampled configurations with negative signed band gaps at temperature $T$.
Figure~\ref{fig-band-inversion-prob} shows the results for \ac{AIMD} and \ac{MCS}.
Both methods show that all sampled configurations remain inverted up to 600~K, whereas at higher temperatures an increasing fraction crosses into normal band ordering.
We therefore conclude that the band inversion associated with the topological phase of \ce{Bi2Se3} remains robust up to $\sim600$~K.
This conclusion is consistent with temperature-dependent \ac{ARPES} measurements, which observe the topological surface state of \ce{Bi2Se3} up to 400~K \cite{Hatch_PRB_83_2011}, as well as spin-resolved \ac{ARPES} measurements demonstrating that the bulk gap remains inverted at room temperature \cite{Pan_PRB_88_2013}.

\subsection{Limitations}

The finite $3\times3\times1$ supercell includes the zone-center phonon modes but does not sample small nonzero wave vectors near $\Gamma$, and therefore misses most of the long-wavelength Fr\"ohlich contribution from polar optical modes \cite{Nery_PRB_94_2016, Chen_PRB_109_2024}.
This contribution can be estimated using Eq.~(2) of \citet{Nery_PRB_94_2016}.
\ce{Bi2Se3} has two infrared-active in-plane polar modes, for which \citet{Koehler_PSSB_61_1974} report the transverse frequencies and oscillator strengths together with fitted in-plane static and high-frequency dielectric constants of $\epsilon_0=96$ and $\epsilon_{\infty}=29$, respectively.
These oscillator parameters give the corresponding longitudinal energies of $\hbar\omega_\mathrm{LO}=10$ and 18~meV.
Decomposition of the inverse dielectric response shows that the upper mode carries approximately 95\% of the long-range Fr\"ohlich coupling weight.
Both modes are included in the estimate.
Effective masses of $m^*_\mathrm{c}=0.124\,m_0$ \cite{Koehler_PSSB_58_1973} and $m^*_\mathrm{v}=0.245\pm0.015\,m_0$ \cite{Piot_PRB_93_2016} are used.
The wave-vector cutoff of $q_F=0.25$~\AA$^{-1}$ is taken as the radius of a sphere whose volume equals that of the supercell \ac{BZ}, so that it spans the wave vectors the supercell cannot represent.
With this the Fr\"ohlich contribution reduces the magnitude of the band gap by 9~meV at 0~K and by 30~meV at 600~K.
The result is not sensitive to this choice, since removing the cutoff altogether increases the correction by about 10\%.
Since only in-plane optical experimental data are available, out-of-plane contributions ($\bm{E}\parallel\bm{c}$) are not evaluated.

Another limitation is the use of a ground-state exchange-correlation functional, which lacks an explicit electronic-temperature dependence.
For zinc-blende semimetals, \citet{Trushin_PRL_120_2018} demonstrated temperature-induced changes in band dispersion and electronic phase transitions using a temperature-dependent exact-exchange approach.
We could not identify an available implementation matching that approach and therefore assessed this effect in \ce{Bi2Se3} using the KDT16 finite-temperature generalized-gradient exchange-correlation functional \citep{Karasiev_PRL_120_2018}, as implemented in ABINIT (v10.8.3) \citep{Verstraete_JCP_163_2025}.
These calculations employed the \ac{PAW} method \citep{Torrent_CMS_42_2008}, included \ac{SOC}, and kept the crystal structure fixed to the VASP equilibrium lattice parameters at 0~K.
\Ac{PBE} frozen-core \ac{PAW} datasets were used with a plane-wave cutoff of 15~Ha in the rhombohedral primitive cell.
The direct band gap at $\Gamma$ increased by only 0.3~meV as the temperature increased from near-zero to 1000~K.
Within this approximation, the electronic-temperature contribution to the gap is negligible on the energy scale considered here.

\textbf{Note:} An earlier version of the present work, including the ensemble-statistical treatment of finite-temperature band inversion and snapshot-by-snapshot \ac{SOC}-scaling analysis, was presented at the 21st International Conference on Density Functional Theory and its Applications \cite{Rubel_DFT2026}.
While finalizing this manuscript, we became aware of an independent study by \citet{Lu_N_x_2026} that shares several methodological elements with the present work.
They performed $NpT$ \ac{AIMD} for the same 135-atom $3\times3\times1$ \ce{Bi2Se3} supercell at the \ac{PBE}-D3 level with Becke-Johnson damping, collected snapshots every 50~fs and analyzed the temperature dependence of the band-gap distribution and of the fraction of snapshots with normal band ordering.
Their band gap is defined at $\Gamma$, and its sign in each snapshot is assigned from the atomic character of the band-edge states and verified by $\mathbb{Z}_2$ calculations for a subset of snapshots.
They observed snapshots with normal band ordering from 500~K and attributed the transition to a reduction of the \ac{SOC}-induced band-gap correction with increasing temperature.
In our results, emergence of snapshots with normal band ordering starts above 600~K.
The difference could be attributed to various snapshot sampling strategies as well as $NpT$ vs $NVT$ ensemble simulation.
Their work did not include a comparison of the \ac{AIMD} ensemble with designed harmonic representations of thermal disorder.
They also did not explore the minimum band gap in the vicinity of $\Gamma$.

\section{Conclusions}

The electronic and topological response of the thermally averaged structure is not equivalent to the ensemble of responses of instantaneous thermally disordered structures.
The thermally averaged structure misses the effects of instantaneous symmetry breaking and shows the least resemblance to the \ac{AIMD} ensemble, underestimating the band gap renormalization with temperature by a large margin.
In particular, inversion-symmetry breaking in thermally distorted structures produces Rashba-like spin splitting away from $\Gamma$, which survives in the ensemble electronic spectrum even though it is completely absent from the band structure of the thermally averaged structure.
Designed thermally disordered structures provide a substantially better representation of the \ac{AIMD} ensemble.
Among them, \ac{MCS} gives the closest agreement with \ac{AIMD}, including the thermal broadening of the band gap, whereas the single \ac{ZG} structure generally overestimates the thermal effect.

\Ac{SOC} scaling offers a convenient high-throughput classification of the band inversion in disordered supercells, provided the $k$ mesh is dense enough to include a generic point where the crossover takes place.
Evaluating the gap only at $\Gamma$ overestimates its magnitude in distorted configurations, since the minimum gap generally lies away from $\Gamma$.

The initially inverted band gap of \ce{Bi2Se3} progressively decreases in magnitude and exhibits increasing thermal broadening with temperature.
To quantify the resulting fluctuations in band ordering, we introduced the band-inversion survival probability, corresponding to the fraction of configurations with negative signed band gaps.
Unlike the mean band gap, the band-inversion survival probability resolves the tail of the band-gap distribution, which determines when individual configurations cross into normal band ordering.
Both \ac{AIMD} and \ac{MCS} show that all sampled configurations remain inverted up to 600~K, whereas at higher temperatures an increasing fraction crosses into normal band ordering.
We therefore conclude that the band inversion associated with the topological phase of \ce{Bi2Se3} remains robust up to approximately 600~K.
More generally, whenever thermal fluctuations bring individual configurations close to an electronic or topological crossover, the finite-temperature properties are set by the distribution over configurations, which no single representative structure, including the thermally averaged one, can be expected to reproduce.

\begin{acknowledgments}
The authors acknowledge funding provided by the Natural Sciences and Engineering Research Council of Canada under the Discovery Grant Programs RGPIN-2025-054113. Computing resources were provided by the Digital Research Alliance of Canada. The authors used Claude (Anthropic) for auditing purposes and cross-checking of data, and take full responsibility for the content of the manuscript.
\end{acknowledgments}

\section*{Data availability}

The data that support the findings of this article are openly available on Zenodo repository (DOI: 10.5281/zenodo.23067045).

\bibliography{bibliography}

\begin{thebibliography}{49}%
\makeatletter
\providecommand \@ifxundefined [1]{%
 \@ifx{#1\undefined}
}%
\providecommand \@ifnum [1]{%
 \ifnum #1\expandafter \@firstoftwo
 \else \expandafter \@secondoftwo
 \fi
}%
\providecommand \@ifx [1]{%
 \ifx #1\expandafter \@firstoftwo
 \else \expandafter \@secondoftwo
 \fi
}%
\providecommand \natexlab [1]{#1}%
\providecommand \enquote  [1]{``#1''}%
\providecommand \bibnamefont  [1]{#1}%
\providecommand \bibfnamefont [1]{#1}%
\providecommand \citenamefont [1]{#1}%
\providecommand \href@noop [0]{\@secondoftwo}%
\providecommand \href [0]{\begingroup \@sanitize@url \@href}%
\providecommand \@href[1]{\@@startlink{#1}\@@href}%
\providecommand \@@href[1]{\endgroup#1\@@endlink}%
\providecommand \@sanitize@url [0]{\catcode `\\12\catcode `\$12\catcode `\&12\catcode `\#12\catcode `\^12\catcode `\_12\catcode `\%12\relax}%
\providecommand \@@startlink[1]{}%
\providecommand \@@endlink[0]{}%
\providecommand \url  [0]{\begingroup\@sanitize@url \@url }%
\providecommand \@url [1]{\endgroup\@href {#1}{\urlprefix }}%
\providecommand \urlprefix  [0]{URL }%
\providecommand \Eprint [0]{\href }%
\providecommand \doibase [0]{https://doi.org/}%
\providecommand \selectlanguage [0]{\@gobble}%
\providecommand \bibinfo  [0]{\@secondoftwo}%
\providecommand \bibfield  [0]{\@secondoftwo}%
\providecommand \translation [1]{[#1]}%
\providecommand \BibitemOpen [0]{}%
\providecommand \bibitemStop [0]{}%
\providecommand \bibitemNoStop [0]{.\EOS\space}%
\providecommand \EOS [0]{\spacefactor3000\relax}%
\providecommand \BibitemShut  [1]{\csname bibitem#1\endcsname}%
\let\auto@bib@innerbib\@empty
\bibitem [{\citenamefont {Allen}\ and\ \citenamefont {Heine}(1976)}]{Allen_JPCSSP_9_1976}%
  \BibitemOpen
  \bibfield  {author} {\bibinfo {author} {\bibfnamefont {P.~B.}\ \bibnamefont {Allen}}\ and\ \bibinfo {author} {\bibfnamefont {V.}~\bibnamefont {Heine}},\ }\bibfield  {title} {\bibinfo {title} {Theory of the temperature dependence of electronic band structures},\ }\href {https://doi.org/10.1088/0022-3719/9/12/013} {\bibfield  {journal} {\bibinfo  {journal} {J. Phys. C: Solid State Phys.}\ }\textbf {\bibinfo {volume} {9}},\ \bibinfo {pages} {2305} (\bibinfo {year} {1976})}\BibitemShut {NoStop}%
\bibitem [{\citenamefont {Giustino}(2017)}]{Giustino_RMP_89_2017}%
  \BibitemOpen
  \bibfield  {author} {\bibinfo {author} {\bibfnamefont {F.}~\bibnamefont {Giustino}},\ }\bibfield  {title} {\bibinfo {title} {Electron-phonon interactions from first principles},\ }\href {https://doi.org/10.1103/RevModPhys.89.015003} {\bibfield  {journal} {\bibinfo  {journal} {Rev. Mod. Phys.}\ }\textbf {\bibinfo {volume} {89}},\ \bibinfo {pages} {015003} (\bibinfo {year} {2017})}\BibitemShut {NoStop}%
\bibitem [{\citenamefont {Garate}(2013)}]{Garate_PRL_110_2013}%
  \BibitemOpen
  \bibfield  {author} {\bibinfo {author} {\bibfnamefont {I.}~\bibnamefont {Garate}},\ }\bibfield  {title} {\bibinfo {title} {Phonon-induced topological transitions and crossovers in {D}irac materials},\ }\href {https://doi.org/10.1103/PhysRevLett.110.046402} {\bibfield  {journal} {\bibinfo  {journal} {Phys. Rev. Lett.}\ }\textbf {\bibinfo {volume} {110}},\ \bibinfo {pages} {046402} (\bibinfo {year} {2013})}\BibitemShut {NoStop}%
\bibitem [{\citenamefont {Kim}\ and\ \citenamefont {Jhi}(2015)}]{Kim_PRB_92_2015}%
  \BibitemOpen
  \bibfield  {author} {\bibinfo {author} {\bibfnamefont {J.}~\bibnamefont {Kim}}\ and\ \bibinfo {author} {\bibfnamefont {S.-H.}\ \bibnamefont {Jhi}},\ }\bibfield  {title} {\bibinfo {title} {Topological phase transitions in group {IV--VI} semiconductors by phonons},\ }\href {https://doi.org/10.1103/PhysRevB.92.125142} {\bibfield  {journal} {\bibinfo  {journal} {Phys. Rev. B}\ }\textbf {\bibinfo {volume} {92}},\ \bibinfo {pages} {125142} (\bibinfo {year} {2015})}\BibitemShut {NoStop}%
\bibitem [{\citenamefont {Antonius}\ and\ \citenamefont {Louie}(2016)}]{Antonius_PRL_117_2016}%
  \BibitemOpen
  \bibfield  {author} {\bibinfo {author} {\bibfnamefont {G.}~\bibnamefont {Antonius}}\ and\ \bibinfo {author} {\bibfnamefont {S.~G.}\ \bibnamefont {Louie}},\ }\bibfield  {title} {\bibinfo {title} {Temperature-induced topological phase transitions: {P}romoted versus suppressed nontrivial topology},\ }\href {https://doi.org/10.1103/PhysRevLett.117.246401} {\bibfield  {journal} {\bibinfo  {journal} {Phys. Rev. Lett.}\ }\textbf {\bibinfo {volume} {117}},\ \bibinfo {pages} {246401} (\bibinfo {year} {2016})}\BibitemShut {NoStop}%
\bibitem [{\citenamefont {Monserrat}\ and\ \citenamefont {Vanderbilt}(2016)}]{Monserrat_PRL_117_2016}%
  \BibitemOpen
  \bibfield  {author} {\bibinfo {author} {\bibfnamefont {B.}~\bibnamefont {Monserrat}}\ and\ \bibinfo {author} {\bibfnamefont {D.}~\bibnamefont {Vanderbilt}},\ }\bibfield  {title} {\bibinfo {title} {Temperature effects in the band structure of topological insulators},\ }\href {https://doi.org/10.1103/PhysRevLett.117.226801} {\bibfield  {journal} {\bibinfo  {journal} {Phys. Rev. Lett.}\ }\textbf {\bibinfo {volume} {117}},\ \bibinfo {pages} {226801} (\bibinfo {year} {2016})}\BibitemShut {NoStop}%
\bibitem [{\citenamefont {Focassio}\ \emph {et~al.}(2021)\citenamefont {Focassio}, \citenamefont {Schleder}, \citenamefont {Crasto~de Lima}, \citenamefont {Lewenkopf},\ and\ \citenamefont {Fazzio}}]{Focassio_PRB_104_2021}%
  \BibitemOpen
  \bibfield  {author} {\bibinfo {author} {\bibfnamefont {B.}~\bibnamefont {Focassio}}, \bibinfo {author} {\bibfnamefont {G.~R.}\ \bibnamefont {Schleder}}, \bibinfo {author} {\bibfnamefont {F.}~\bibnamefont {Crasto~de Lima}}, \bibinfo {author} {\bibfnamefont {C.}~\bibnamefont {Lewenkopf}},\ and\ \bibinfo {author} {\bibfnamefont {A.}~\bibnamefont {Fazzio}},\ }\bibfield  {title} {\bibinfo {title} {Amorphous {Bi$_2$Se$_3$} structural, electronic, and topological nature from first principles},\ }\href {https://doi.org/10.1103/PhysRevB.104.214206} {\bibfield  {journal} {\bibinfo  {journal} {Phys. Rev. B}\ }\textbf {\bibinfo {volume} {104}},\ \bibinfo {pages} {214206} (\bibinfo {year} {2021})}\BibitemShut {NoStop}%
\bibitem [{\citenamefont {Corbae}\ \emph {et~al.}(2023)\citenamefont {Corbae}, \citenamefont {Ciocys}, \citenamefont {Varjas}, \citenamefont {Kennedy}, \citenamefont {Zeltmann}, \citenamefont {Molina-Ruiz}, \citenamefont {Griffin}, \citenamefont {Jozwiak}, \citenamefont {Chen}, \citenamefont {Wang}, \citenamefont {Minor}, \citenamefont {Scott}, \citenamefont {Grushin}, \citenamefont {Lanzara},\ and\ \citenamefont {Hellman}}]{Corbae_NM_22_2023}%
  \BibitemOpen
  \bibfield  {author} {\bibinfo {author} {\bibfnamefont {P.}~\bibnamefont {Corbae}}, \bibinfo {author} {\bibfnamefont {S.}~\bibnamefont {Ciocys}}, \bibinfo {author} {\bibfnamefont {D.}~\bibnamefont {Varjas}}, \bibinfo {author} {\bibfnamefont {E.}~\bibnamefont {Kennedy}}, \bibinfo {author} {\bibfnamefont {S.}~\bibnamefont {Zeltmann}}, \bibinfo {author} {\bibfnamefont {M.}~\bibnamefont {Molina-Ruiz}}, \bibinfo {author} {\bibfnamefont {S.~M.}\ \bibnamefont {Griffin}}, \bibinfo {author} {\bibfnamefont {C.}~\bibnamefont {Jozwiak}}, \bibinfo {author} {\bibfnamefont {Z.}~\bibnamefont {Chen}}, \bibinfo {author} {\bibfnamefont {L.-W.}\ \bibnamefont {Wang}}, \bibinfo {author} {\bibfnamefont {A.~M.}\ \bibnamefont {Minor}}, \bibinfo {author} {\bibfnamefont {M.}~\bibnamefont {Scott}}, \bibinfo {author} {\bibfnamefont {A.~G.}\ \bibnamefont {Grushin}}, \bibinfo {author} {\bibfnamefont {A.}~\bibnamefont {Lanzara}},\ and\ \bibinfo {author} {\bibfnamefont {F.}~\bibnamefont {Hellman}},\ }\bibfield  {title} {\bibinfo {title}
  {Observation of spin-momentum locked surface states in amorphous {Bi$_2$Se$_3$}},\ }\href {https://doi.org/10.1038/s41563-022-01458-0} {\bibfield  {journal} {\bibinfo  {journal} {Nat. Mater.}\ }\textbf {\bibinfo {volume} {22}},\ \bibinfo {pages} {200} (\bibinfo {year} {2023})}\BibitemShut {NoStop}%
\bibitem [{\citenamefont {Zacharias}\ \emph {et~al.}(2020)\citenamefont {Zacharias}, \citenamefont {Scheffler},\ and\ \citenamefont {Carbogno}}]{Zacharias_PRB_102_2020}%
  \BibitemOpen
  \bibfield  {author} {\bibinfo {author} {\bibfnamefont {M.}~\bibnamefont {Zacharias}}, \bibinfo {author} {\bibfnamefont {M.}~\bibnamefont {Scheffler}},\ and\ \bibinfo {author} {\bibfnamefont {C.}~\bibnamefont {Carbogno}},\ }\bibfield  {title} {\bibinfo {title} {Fully anharmonic nonperturbative theory of vibronically renormalized electronic band structures},\ }\href {https://doi.org/10.1103/PhysRevB.102.045126} {\bibfield  {journal} {\bibinfo  {journal} {Phys. Rev. B}\ }\textbf {\bibinfo {volume} {102}},\ \bibinfo {pages} {045126} (\bibinfo {year} {2020})}\BibitemShut {NoStop}%
\bibitem [{\citenamefont {Kim}\ \emph {et~al.}(2012)\citenamefont {Kim}, \citenamefont {Chen}, \citenamefont {Wang}, \citenamefont {Shi}, \citenamefont {Miotkowski}, \citenamefont {Chen}, \citenamefont {Sharma}, \citenamefont {Lima~Sharma}, \citenamefont {Hekmaty}, \citenamefont {Jiang},\ and\ \citenamefont {Smirnov}}]{Kim_APL_100_2012}%
  \BibitemOpen
  \bibfield  {author} {\bibinfo {author} {\bibfnamefont {Y.}~\bibnamefont {Kim}}, \bibinfo {author} {\bibfnamefont {X.}~\bibnamefont {Chen}}, \bibinfo {author} {\bibfnamefont {Z.}~\bibnamefont {Wang}}, \bibinfo {author} {\bibfnamefont {J.}~\bibnamefont {Shi}}, \bibinfo {author} {\bibfnamefont {I.}~\bibnamefont {Miotkowski}}, \bibinfo {author} {\bibfnamefont {Y.~P.}\ \bibnamefont {Chen}}, \bibinfo {author} {\bibfnamefont {P.~A.}\ \bibnamefont {Sharma}}, \bibinfo {author} {\bibfnamefont {A.~L.}\ \bibnamefont {Lima~Sharma}}, \bibinfo {author} {\bibfnamefont {M.~A.}\ \bibnamefont {Hekmaty}}, \bibinfo {author} {\bibfnamefont {Z.}~\bibnamefont {Jiang}},\ and\ \bibinfo {author} {\bibfnamefont {D.}~\bibnamefont {Smirnov}},\ }\bibfield  {title} {\bibinfo {title} {Temperature dependence of {R}aman-active optical phonons in {Bi$_2$Se$_3$} and {Sb$_2$Te$_3$}},\ }\href {https://doi.org/10.1063/1.3685465} {\bibfield  {journal} {\bibinfo  {journal} {Appl. Phys. Lett.}\ }\textbf {\bibinfo {volume} {100}},\ \bibinfo {pages}
  {071907} (\bibinfo {year} {2012})}\BibitemShut {NoStop}%
\bibitem [{\citenamefont {Tian}\ \emph {et~al.}(2017)\citenamefont {Tian}, \citenamefont {Jia}, \citenamefont {Cava}, \citenamefont {Zhong}, \citenamefont {Schneeloch}, \citenamefont {Gu},\ and\ \citenamefont {Burch}}]{Tian_PRB_95_2017}%
  \BibitemOpen
  \bibfield  {author} {\bibinfo {author} {\bibfnamefont {Y.}~\bibnamefont {Tian}}, \bibinfo {author} {\bibfnamefont {S.}~\bibnamefont {Jia}}, \bibinfo {author} {\bibfnamefont {R.~J.}\ \bibnamefont {Cava}}, \bibinfo {author} {\bibfnamefont {R.}~\bibnamefont {Zhong}}, \bibinfo {author} {\bibfnamefont {J.}~\bibnamefont {Schneeloch}}, \bibinfo {author} {\bibfnamefont {G.}~\bibnamefont {Gu}},\ and\ \bibinfo {author} {\bibfnamefont {K.~S.}\ \bibnamefont {Burch}},\ }\bibfield  {title} {\bibinfo {title} {Understanding the evolution of anomalous anharmonicity in {Bi$_2$Te$_{3-x}$Se$_x$}},\ }\href {https://doi.org/10.1103/physrevb.95.094104} {\bibfield  {journal} {\bibinfo  {journal} {Phys. Rev. B}\ }\textbf {\bibinfo {volume} {95}},\ \bibinfo {pages} {094104} (\bibinfo {year} {2017})}\BibitemShut {NoStop}%
\bibitem [{\citenamefont {Park}\ \emph {et~al.}(2013)\citenamefont {Park}, \citenamefont {Nomura}, \citenamefont {Arita}, \citenamefont {Llobet},\ and\ \citenamefont {Louca}}]{Park_PRB_88_2013}%
  \BibitemOpen
  \bibfield  {author} {\bibinfo {author} {\bibfnamefont {K.}~\bibnamefont {Park}}, \bibinfo {author} {\bibfnamefont {Y.}~\bibnamefont {Nomura}}, \bibinfo {author} {\bibfnamefont {R.}~\bibnamefont {Arita}}, \bibinfo {author} {\bibfnamefont {A.}~\bibnamefont {Llobet}},\ and\ \bibinfo {author} {\bibfnamefont {D.}~\bibnamefont {Louca}},\ }\bibfield  {title} {\bibinfo {title} {Local strain and anharmonicity in the bonding of {Bi$_2$Se$_{3-x}$Te$_x$} topological insulators},\ }\href {https://doi.org/10.1103/physrevb.88.224108} {\bibfield  {journal} {\bibinfo  {journal} {Phys. Rev. B}\ }\textbf {\bibinfo {volume} {88}},\ \bibinfo {pages} {224108} (\bibinfo {year} {2013})}\BibitemShut {NoStop}%
\bibitem [{\citenamefont {Kresse}\ and\ \citenamefont {Furthm{\"u}ller}(1996)}]{Kresse_PRB_54_1996}%
  \BibitemOpen
  \bibfield  {author} {\bibinfo {author} {\bibfnamefont {G.}~\bibnamefont {Kresse}}\ and\ \bibinfo {author} {\bibfnamefont {J.}~\bibnamefont {Furthm{\"u}ller}},\ }\bibfield  {title} {\bibinfo {title} {Efficient iterative schemes for ab initio total-energy calculations using a plane-wave basis set},\ }\href {https://doi.org/10.1103/PhysRevB.54.11169} {\bibfield  {journal} {\bibinfo  {journal} {Phys. Rev. B}\ }\textbf {\bibinfo {volume} {54}},\ \bibinfo {pages} {11169} (\bibinfo {year} {1996})}\BibitemShut {NoStop}%
\bibitem [{\citenamefont {Kresse}\ and\ \citenamefont {Joubert}(1999)}]{Kresse_PRB_59_1999}%
  \BibitemOpen
  \bibfield  {author} {\bibinfo {author} {\bibfnamefont {G.}~\bibnamefont {Kresse}}\ and\ \bibinfo {author} {\bibfnamefont {D.}~\bibnamefont {Joubert}},\ }\bibfield  {title} {\bibinfo {title} {From ultrasoft pseudopotentials to the projector augmented-wave method},\ }\href {https://doi.org/10.1103/PhysRevB.59.1758} {\bibfield  {journal} {\bibinfo  {journal} {Phys. Rev. B}\ }\textbf {\bibinfo {volume} {59}},\ \bibinfo {pages} {1758} (\bibinfo {year} {1999})}\BibitemShut {NoStop}%
\bibitem [{\citenamefont {Perdew}\ \emph {et~al.}(1996)\citenamefont {Perdew}, \citenamefont {Burke},\ and\ \citenamefont {Ernzerhof}}]{Perdew_PRL_77_1996}%
  \BibitemOpen
  \bibfield  {author} {\bibinfo {author} {\bibfnamefont {J.~P.}\ \bibnamefont {Perdew}}, \bibinfo {author} {\bibfnamefont {K.}~\bibnamefont {Burke}},\ and\ \bibinfo {author} {\bibfnamefont {M.}~\bibnamefont {Ernzerhof}},\ }\bibfield  {title} {\bibinfo {title} {Generalized gradient approximation made simple},\ }\href {https://doi.org/10.1103/PhysRevLett.77.3865} {\bibfield  {journal} {\bibinfo  {journal} {Phys. Rev. Lett.}\ }\textbf {\bibinfo {volume} {77}},\ \bibinfo {pages} {3865} (\bibinfo {year} {1996})}\BibitemShut {NoStop}%
\bibitem [{\citenamefont {Grimme}\ \emph {et~al.}(2010)\citenamefont {Grimme}, \citenamefont {Antony}, \citenamefont {Ehrlich},\ and\ \citenamefont {Krieg}}]{Grimme_JCP_132_2010}%
  \BibitemOpen
  \bibfield  {author} {\bibinfo {author} {\bibfnamefont {S.}~\bibnamefont {Grimme}}, \bibinfo {author} {\bibfnamefont {J.}~\bibnamefont {Antony}}, \bibinfo {author} {\bibfnamefont {S.}~\bibnamefont {Ehrlich}},\ and\ \bibinfo {author} {\bibfnamefont {H.}~\bibnamefont {Krieg}},\ }\bibfield  {title} {\bibinfo {title} {A consistent and accurate ab initio parametrization of density functional dispersion correction ({DFT-D}) for the 94 elements {H-Pu}},\ }\href {https://doi.org/10.1063/1.3382344} {\bibfield  {journal} {\bibinfo  {journal} {J. Chem. Phys.}\ }\textbf {\bibinfo {volume} {132}},\ \bibinfo {pages} {154104} (\bibinfo {year} {2010})}\BibitemShut {NoStop}%
\bibitem [{\citenamefont {Zacharias}\ and\ \citenamefont {Giustino}(2016)}]{Zacharias_PRB_94_2016}%
  \BibitemOpen
  \bibfield  {author} {\bibinfo {author} {\bibfnamefont {M.}~\bibnamefont {Zacharias}}\ and\ \bibinfo {author} {\bibfnamefont {F.}~\bibnamefont {Giustino}},\ }\bibfield  {title} {\bibinfo {title} {One-shot calculation of temperature-dependent optical spectra and phonon-induced band-gap renormalization},\ }\href {https://doi.org/10.1103/physrevb.94.075125} {\bibfield  {journal} {\bibinfo  {journal} {Phys. Rev. B}\ }\textbf {\bibinfo {volume} {94}},\ \bibinfo {pages} {075125} (\bibinfo {year} {2016})}\BibitemShut {NoStop}%
\bibitem [{\citenamefont {Karsai}\ \emph {et~al.}(2018)\citenamefont {Karsai}, \citenamefont {Engel}, \citenamefont {Flage-Larsen},\ and\ \citenamefont {Kresse}}]{Karsai_NJP_20_2018}%
  \BibitemOpen
  \bibfield  {author} {\bibinfo {author} {\bibfnamefont {F.}~\bibnamefont {Karsai}}, \bibinfo {author} {\bibfnamefont {M.}~\bibnamefont {Engel}}, \bibinfo {author} {\bibfnamefont {E.}~\bibnamefont {Flage-Larsen}},\ and\ \bibinfo {author} {\bibfnamefont {G.}~\bibnamefont {Kresse}},\ }\bibfield  {title} {\bibinfo {title} {Electron--phonon coupling in semiconductors within the {GW} approximation},\ }\href {https://doi.org/10.1088/1367-2630/aaf53f} {\bibfield  {journal} {\bibinfo  {journal} {New J. Phys.}\ }\textbf {\bibinfo {volume} {20}},\ \bibinfo {pages} {123008} (\bibinfo {year} {2018})}\BibitemShut {NoStop}%
\bibitem [{\citenamefont {Parrinello}\ and\ \citenamefont {Rahman}(1980)}]{Parrinello_PRL_45_1980}%
  \BibitemOpen
  \bibfield  {author} {\bibinfo {author} {\bibfnamefont {M.}~\bibnamefont {Parrinello}}\ and\ \bibinfo {author} {\bibfnamefont {A.}~\bibnamefont {Rahman}},\ }\bibfield  {title} {\bibinfo {title} {Crystal structure and pair potentials: {A} molecular-dynamics study},\ }\href {https://doi.org/10.1103/physrevlett.45.1196} {\bibfield  {journal} {\bibinfo  {journal} {Phys. Rev. Lett.}\ }\textbf {\bibinfo {volume} {45}},\ \bibinfo {pages} {1196} (\bibinfo {year} {1980})}\BibitemShut {NoStop}%
\bibitem [{\citenamefont {Parrinello}\ and\ \citenamefont {Rahman}(1981)}]{Parrinello_JAP_52_1981}%
  \BibitemOpen
  \bibfield  {author} {\bibinfo {author} {\bibfnamefont {M.}~\bibnamefont {Parrinello}}\ and\ \bibinfo {author} {\bibfnamefont {A.}~\bibnamefont {Rahman}},\ }\bibfield  {title} {\bibinfo {title} {Polymorphic transitions in single crystals: {A} new molecular dynamics method},\ }\href {https://doi.org/10.1063/1.328693} {\bibfield  {journal} {\bibinfo  {journal} {J. Appl. Phys.}\ }\textbf {\bibinfo {volume} {52}},\ \bibinfo {pages} {7182} (\bibinfo {year} {1981})}\BibitemShut {NoStop}%
\bibitem [{\citenamefont {Jinnouchi}\ \emph {et~al.}(2019)\citenamefont {Jinnouchi}, \citenamefont {Karsai},\ and\ \citenamefont {Kresse}}]{Jinnouchi_PRB_100_2019}%
  \BibitemOpen
  \bibfield  {author} {\bibinfo {author} {\bibfnamefont {R.}~\bibnamefont {Jinnouchi}}, \bibinfo {author} {\bibfnamefont {F.}~\bibnamefont {Karsai}},\ and\ \bibinfo {author} {\bibfnamefont {G.}~\bibnamefont {Kresse}},\ }\bibfield  {title} {\bibinfo {title} {On-the-fly machine learning force field generation: {A}pplication to melting points},\ }\href {https://doi.org/10.1103/physrevb.100.014105} {\bibfield  {journal} {\bibinfo  {journal} {Phys. Rev. B}\ }\textbf {\bibinfo {volume} {100}},\ \bibinfo {pages} {014105} (\bibinfo {year} {2019})}\BibitemShut {NoStop}%
\bibitem [{\citenamefont {Bussi}\ \emph {et~al.}(2007)\citenamefont {Bussi}, \citenamefont {Donadio},\ and\ \citenamefont {Parrinello}}]{Bussi_JCP_126_2007}%
  \BibitemOpen
  \bibfield  {author} {\bibinfo {author} {\bibfnamefont {G.}~\bibnamefont {Bussi}}, \bibinfo {author} {\bibfnamefont {D.}~\bibnamefont {Donadio}},\ and\ \bibinfo {author} {\bibfnamefont {M.}~\bibnamefont {Parrinello}},\ }\bibfield  {title} {\bibinfo {title} {Canonical sampling through velocity rescaling},\ }\href {https://doi.org/10.1063/1.2408420} {\bibfield  {journal} {\bibinfo  {journal} {J. Chem. Phys.}\ }\textbf {\bibinfo {volume} {126}},\ \bibinfo {pages} {014101} (\bibinfo {year} {2007})}\BibitemShut {NoStop}%
\bibitem [{\citenamefont {Wang}\ and\ \citenamefont {Johnson}(2011)}]{Wang_PRB_83_2011}%
  \BibitemOpen
  \bibfield  {author} {\bibinfo {author} {\bibfnamefont {L.-L.}\ \bibnamefont {Wang}}\ and\ \bibinfo {author} {\bibfnamefont {D.~D.}\ \bibnamefont {Johnson}},\ }\bibfield  {title} {\bibinfo {title} {Ternary tetradymite compounds as topological insulators},\ }\href {https://doi.org/10.1103/PhysRevB.83.241309} {\bibfield  {journal} {\bibinfo  {journal} {Phys. Rev. B}\ }\textbf {\bibinfo {volume} {83}},\ \bibinfo {pages} {241309(R)} (\bibinfo {year} {2011})}\BibitemShut {NoStop}%
\bibitem [{\citenamefont {Gomez-Bastidas}\ and\ \citenamefont {Rubel}(2023)}]{GomezBastidas_CPC_292_2023}%
  \BibitemOpen
  \bibfield  {author} {\bibinfo {author} {\bibfnamefont {A.~F.}\ \bibnamefont {Gomez-Bastidas}}\ and\ \bibinfo {author} {\bibfnamefont {O.}~\bibnamefont {Rubel}},\ }\bibfield  {title} {\bibinfo {title} {Software implementation for calculating {C}hern and {$\mathbb{Z}_2$} topological invariants of crystalline solids with {WIEN2k} all-electron density functional package},\ }\href {https://doi.org/10.1016/j.cpc.2023.108864} {\bibfield  {journal} {\bibinfo  {journal} {Comput. Phys. Commun.}\ }\textbf {\bibinfo {volume} {292}},\ \bibinfo {pages} {108864} (\bibinfo {year} {2023})}\BibitemShut {NoStop}%
\bibitem [{\citenamefont {Chen}\ \emph {et~al.}(2011)\citenamefont {Chen}, \citenamefont {Zhou}, \citenamefont {Kiswandhi}, \citenamefont {Miotkowski}, \citenamefont {Chen}, \citenamefont {Sharma}, \citenamefont {Lima~Sharma}, \citenamefont {Hekmaty}, \citenamefont {Smirnov},\ and\ \citenamefont {Jiang}}]{Chen_APL_99_2011}%
  \BibitemOpen
  \bibfield  {author} {\bibinfo {author} {\bibfnamefont {X.}~\bibnamefont {Chen}}, \bibinfo {author} {\bibfnamefont {H.~D.}\ \bibnamefont {Zhou}}, \bibinfo {author} {\bibfnamefont {A.}~\bibnamefont {Kiswandhi}}, \bibinfo {author} {\bibfnamefont {I.}~\bibnamefont {Miotkowski}}, \bibinfo {author} {\bibfnamefont {Y.~P.}\ \bibnamefont {Chen}}, \bibinfo {author} {\bibfnamefont {P.~A.}\ \bibnamefont {Sharma}}, \bibinfo {author} {\bibfnamefont {A.~L.}\ \bibnamefont {Lima~Sharma}}, \bibinfo {author} {\bibfnamefont {M.~A.}\ \bibnamefont {Hekmaty}}, \bibinfo {author} {\bibfnamefont {D.}~\bibnamefont {Smirnov}},\ and\ \bibinfo {author} {\bibfnamefont {Z.}~\bibnamefont {Jiang}},\ }\bibfield  {title} {\bibinfo {title} {Thermal expansion coefficients of {Bi$_2$Se$_3$} and {Sb$_2$Te$_3$} crystals from 10~{K} to 270~{K}},\ }\href {https://doi.org/10.1063/1.3672198} {\bibfield  {journal} {\bibinfo  {journal} {Appl. Phys. Lett.}\ }\textbf {\bibinfo {volume} {99}},\ \bibinfo {pages} {261912} (\bibinfo {year}
  {2011})}\BibitemShut {NoStop}%
\bibitem [{\citenamefont {Kevy}\ \emph {et~al.}(2024)\citenamefont {Kevy}, \citenamefont {Wollesen},\ and\ \citenamefont {Bremholm}}]{Kevy_JSSC_330_2024}%
  \BibitemOpen
  \bibfield  {author} {\bibinfo {author} {\bibfnamefont {S.~M.}\ \bibnamefont {Kevy}}, \bibinfo {author} {\bibfnamefont {L.}~\bibnamefont {Wollesen}},\ and\ \bibinfo {author} {\bibfnamefont {M.}~\bibnamefont {Bremholm}},\ }\bibfield  {title} {\bibinfo {title} {In-situ x-ray diffraction study of {Nb}-doped {Bi$_2$Se$_3$} crystal growth revealing unavoidable misfit layer compound},\ }\href {https://doi.org/10.1016/j.jssc.2023.124477} {\bibfield  {journal} {\bibinfo  {journal} {J. Solid State Chem.}\ }\textbf {\bibinfo {volume} {330}},\ \bibinfo {pages} {124477} (\bibinfo {year} {2024})}\BibitemShut {NoStop}%
\bibitem [{\citenamefont {Nechaev}\ \emph {et~al.}(2013)\citenamefont {Nechaev}, \citenamefont {Hatch}, \citenamefont {Bianchi}, \citenamefont {Guan}, \citenamefont {Friedrich}, \citenamefont {Aguilera}, \citenamefont {Mi}, \citenamefont {Iversen}, \citenamefont {Bl{\"u}gel}, \citenamefont {Hofmann},\ and\ \citenamefont {Chulkov}}]{Nechaev_PRB_87_2013}%
  \BibitemOpen
  \bibfield  {author} {\bibinfo {author} {\bibfnamefont {I.~A.}\ \bibnamefont {Nechaev}}, \bibinfo {author} {\bibfnamefont {R.~C.}\ \bibnamefont {Hatch}}, \bibinfo {author} {\bibfnamefont {M.}~\bibnamefont {Bianchi}}, \bibinfo {author} {\bibfnamefont {D.}~\bibnamefont {Guan}}, \bibinfo {author} {\bibfnamefont {C.}~\bibnamefont {Friedrich}}, \bibinfo {author} {\bibfnamefont {I.}~\bibnamefont {Aguilera}}, \bibinfo {author} {\bibfnamefont {J.~L.}\ \bibnamefont {Mi}}, \bibinfo {author} {\bibfnamefont {B.~B.}\ \bibnamefont {Iversen}}, \bibinfo {author} {\bibfnamefont {S.}~\bibnamefont {Bl{\"u}gel}}, \bibinfo {author} {\bibfnamefont {P.}~\bibnamefont {Hofmann}},\ and\ \bibinfo {author} {\bibfnamefont {E.~V.}\ \bibnamefont {Chulkov}},\ }\bibfield  {title} {\bibinfo {title} {Evidence for a direct band gap in the topological insulator {Bi$_2$Se$_3$} from theory and experiment},\ }\href {https://doi.org/10.1103/PhysRevB.87.121111} {\bibfield  {journal} {\bibinfo  {journal} {Phys. Rev. B}\ }\textbf {\bibinfo {volume}
  {87}},\ \bibinfo {pages} {121111(R)} (\bibinfo {year} {2013})}\BibitemShut {NoStop}%
\bibitem [{\citenamefont {Monserrat}\ and\ \citenamefont {Vanderbilt}(2017)}]{Monserrat_arXiv_1711.06274}%
  \BibitemOpen
  \bibfield  {author} {\bibinfo {author} {\bibfnamefont {B.}~\bibnamefont {Monserrat}}\ and\ \bibinfo {author} {\bibfnamefont {D.}~\bibnamefont {Vanderbilt}},\ }\href@noop {} {\bibinfo {title} {Phonon-assisted spin splitting in centrosymmetric crystals}} (\bibinfo {year} {2017}),\ \Eprint {https://arxiv.org/abs/1711.06274} {arXiv:1711.06274 [cond-mat.mtrl-sci]} \BibitemShut {NoStop}%
\bibitem [{\citenamefont {McKechnie}\ \emph {et~al.}(2018)\citenamefont {McKechnie}, \citenamefont {Frost}, \citenamefont {Pashov}, \citenamefont {Azarhoosh}, \citenamefont {Walsh},\ and\ \citenamefont {van Schilfgaarde}}]{McKechnie_PRB_98_2018}%
  \BibitemOpen
  \bibfield  {author} {\bibinfo {author} {\bibfnamefont {S.}~\bibnamefont {McKechnie}}, \bibinfo {author} {\bibfnamefont {J.~M.}\ \bibnamefont {Frost}}, \bibinfo {author} {\bibfnamefont {D.}~\bibnamefont {Pashov}}, \bibinfo {author} {\bibfnamefont {P.}~\bibnamefont {Azarhoosh}}, \bibinfo {author} {\bibfnamefont {A.}~\bibnamefont {Walsh}},\ and\ \bibinfo {author} {\bibfnamefont {M.}~\bibnamefont {van Schilfgaarde}},\ }\bibfield  {title} {\bibinfo {title} {Dynamic symmetry breaking and spin splitting in metal halide perovskites},\ }\href {https://doi.org/10.1103/PhysRevB.98.085108} {\bibfield  {journal} {\bibinfo  {journal} {Phys. Rev. B}\ }\textbf {\bibinfo {volume} {98}},\ \bibinfo {pages} {085108} (\bibinfo {year} {2018})}\BibitemShut {NoStop}%
\bibitem [{\citenamefont {Niesner}\ \emph {et~al.}(2018)\citenamefont {Niesner}, \citenamefont {Hauck}, \citenamefont {Shrestha}, \citenamefont {Levchuk}, \citenamefont {Matt}, \citenamefont {Osvet}, \citenamefont {Batentschuk}, \citenamefont {Brabec}, \citenamefont {Weber},\ and\ \citenamefont {Fauster}}]{Niesner_PNAS_115_2018}%
  \BibitemOpen
  \bibfield  {author} {\bibinfo {author} {\bibfnamefont {D.}~\bibnamefont {Niesner}}, \bibinfo {author} {\bibfnamefont {M.}~\bibnamefont {Hauck}}, \bibinfo {author} {\bibfnamefont {S.}~\bibnamefont {Shrestha}}, \bibinfo {author} {\bibfnamefont {I.}~\bibnamefont {Levchuk}}, \bibinfo {author} {\bibfnamefont {G.~J.}\ \bibnamefont {Matt}}, \bibinfo {author} {\bibfnamefont {A.}~\bibnamefont {Osvet}}, \bibinfo {author} {\bibfnamefont {M.}~\bibnamefont {Batentschuk}}, \bibinfo {author} {\bibfnamefont {C.}~\bibnamefont {Brabec}}, \bibinfo {author} {\bibfnamefont {H.~B.}\ \bibnamefont {Weber}},\ and\ \bibinfo {author} {\bibfnamefont {T.}~\bibnamefont {Fauster}},\ }\bibfield  {title} {\bibinfo {title} {Structural fluctuations cause spin-split states in tetragonal {(CH$_3$NH$_3$)PbI$_3$} as evidenced by the circular photogalvanic effect},\ }\href {https://doi.org/10.1073/pnas.1805422115} {\bibfield  {journal} {\bibinfo  {journal} {Proc. Natl. Acad. Sci. U.S.A.}\ }\textbf {\bibinfo {volume} {115}},\ \bibinfo {pages}
  {9509} (\bibinfo {year} {2018})}\BibitemShut {NoStop}%
\bibitem [{\citenamefont {Xia}\ \emph {et~al.}(2009)\citenamefont {Xia}, \citenamefont {Qian}, \citenamefont {Hsieh}, \citenamefont {Wray}, \citenamefont {Pal}, \citenamefont {Lin}, \citenamefont {Bansil}, \citenamefont {Grauer}, \citenamefont {Hor}, \citenamefont {Cava},\ and\ \citenamefont {Hasan}}]{Xia_NP_5_2009}%
  \BibitemOpen
  \bibfield  {author} {\bibinfo {author} {\bibfnamefont {Y.}~\bibnamefont {Xia}}, \bibinfo {author} {\bibfnamefont {D.}~\bibnamefont {Qian}}, \bibinfo {author} {\bibfnamefont {D.}~\bibnamefont {Hsieh}}, \bibinfo {author} {\bibfnamefont {L.}~\bibnamefont {Wray}}, \bibinfo {author} {\bibfnamefont {A.}~\bibnamefont {Pal}}, \bibinfo {author} {\bibfnamefont {H.}~\bibnamefont {Lin}}, \bibinfo {author} {\bibfnamefont {A.}~\bibnamefont {Bansil}}, \bibinfo {author} {\bibfnamefont {D.}~\bibnamefont {Grauer}}, \bibinfo {author} {\bibfnamefont {Y.~S.}\ \bibnamefont {Hor}}, \bibinfo {author} {\bibfnamefont {R.~J.}\ \bibnamefont {Cava}},\ and\ \bibinfo {author} {\bibfnamefont {M.~Z.}\ \bibnamefont {Hasan}},\ }\bibfield  {title} {\bibinfo {title} {Observation of a large-gap topological-insulator class with a single {D}irac cone on the surface},\ }\href {https://doi.org/10.1038/nphys1274} {\bibfield  {journal} {\bibinfo  {journal} {Nat. Phys.}\ }\textbf {\bibinfo {volume} {5}},\ \bibinfo {pages} {398} (\bibinfo {year}
  {2009})}\BibitemShut {NoStop}%
\bibitem [{\citenamefont {K{\"o}hler}\ and\ \citenamefont {Hartmann}(1974)}]{Kohler_PSSB_63_1974}%
  \BibitemOpen
  \bibfield  {author} {\bibinfo {author} {\bibfnamefont {H.}~\bibnamefont {K{\"o}hler}}\ and\ \bibinfo {author} {\bibfnamefont {J.}~\bibnamefont {Hartmann}},\ }\bibfield  {title} {\bibinfo {title} {Burstein shift of the absorption edge of n-{Bi$_2$Se$_3$}},\ }\href {https://doi.org/10.1002/pssb.2220630116} {\bibfield  {journal} {\bibinfo  {journal} {Phys. Status Solidi B}\ }\textbf {\bibinfo {volume} {63}},\ \bibinfo {pages} {171} (\bibinfo {year} {1974})}\BibitemShut {NoStop}%
\bibitem [{\citenamefont {Post}\ \emph {et~al.}(2013)\citenamefont {Post}, \citenamefont {Chapler}, \citenamefont {He}, \citenamefont {Kou}, \citenamefont {Wang},\ and\ \citenamefont {Basov}}]{Post_PRB_88_2013}%
  \BibitemOpen
  \bibfield  {author} {\bibinfo {author} {\bibfnamefont {K.~W.}\ \bibnamefont {Post}}, \bibinfo {author} {\bibfnamefont {B.~C.}\ \bibnamefont {Chapler}}, \bibinfo {author} {\bibfnamefont {L.}~\bibnamefont {He}}, \bibinfo {author} {\bibfnamefont {X.}~\bibnamefont {Kou}}, \bibinfo {author} {\bibfnamefont {K.~L.}\ \bibnamefont {Wang}},\ and\ \bibinfo {author} {\bibfnamefont {D.~N.}\ \bibnamefont {Basov}},\ }\bibfield  {title} {\bibinfo {title} {Thickness-dependent bulk electronic properties in {Bi$_2$Se$_3$} thin films revealed by infrared spectroscopy},\ }\href {https://doi.org/10.1103/PhysRevB.88.075121} {\bibfield  {journal} {\bibinfo  {journal} {Phys. Rev. B}\ }\textbf {\bibinfo {volume} {88}},\ \bibinfo {pages} {075121} (\bibinfo {year} {2013})}\BibitemShut {NoStop}%
\bibitem [{\citenamefont {Orlita}\ \emph {et~al.}(2015)\citenamefont {Orlita}, \citenamefont {Piot}, \citenamefont {Martinez}, \citenamefont {Sampath~Kumar}, \citenamefont {Faugeras}, \citenamefont {Potemski}, \citenamefont {Michel}, \citenamefont {Hankiewicz}, \citenamefont {Brauner}, \citenamefont {Dra{\v{s}}ar}, \citenamefont {Schreyeck}, \citenamefont {Grauer}, \citenamefont {Brunner}, \citenamefont {Gould}, \citenamefont {Br{\"u}ne},\ and\ \citenamefont {Molenkamp}}]{Orlita_PRL_114_2015}%
  \BibitemOpen
  \bibfield  {author} {\bibinfo {author} {\bibfnamefont {M.}~\bibnamefont {Orlita}}, \bibinfo {author} {\bibfnamefont {B.~A.}\ \bibnamefont {Piot}}, \bibinfo {author} {\bibfnamefont {G.}~\bibnamefont {Martinez}}, \bibinfo {author} {\bibfnamefont {N.~K.}\ \bibnamefont {Sampath~Kumar}}, \bibinfo {author} {\bibfnamefont {C.}~\bibnamefont {Faugeras}}, \bibinfo {author} {\bibfnamefont {M.}~\bibnamefont {Potemski}}, \bibinfo {author} {\bibfnamefont {C.}~\bibnamefont {Michel}}, \bibinfo {author} {\bibfnamefont {E.~M.}\ \bibnamefont {Hankiewicz}}, \bibinfo {author} {\bibfnamefont {T.}~\bibnamefont {Brauner}}, \bibinfo {author} {\bibfnamefont {{\v{C}}.}~\bibnamefont {Dra{\v{s}}ar}}, \bibinfo {author} {\bibfnamefont {S.}~\bibnamefont {Schreyeck}}, \bibinfo {author} {\bibfnamefont {S.}~\bibnamefont {Grauer}}, \bibinfo {author} {\bibfnamefont {K.}~\bibnamefont {Brunner}}, \bibinfo {author} {\bibfnamefont {C.}~\bibnamefont {Gould}}, \bibinfo {author} {\bibfnamefont {C.}~\bibnamefont {Br{\"u}ne}},\ and\ \bibinfo {author}
  {\bibfnamefont {L.~W.}\ \bibnamefont {Molenkamp}},\ }\bibfield  {title} {\bibinfo {title} {Magneto-optics of massive {Dirac} fermions in bulk {Bi$_2$Se$_3$}},\ }\href {https://doi.org/10.1103/PhysRevLett.114.186401} {\bibfield  {journal} {\bibinfo  {journal} {Phys. Rev. Lett.}\ }\textbf {\bibinfo {volume} {114}},\ \bibinfo {pages} {186401} (\bibinfo {year} {2015})}\BibitemShut {NoStop}%
\bibitem [{\citenamefont {Gobrecht}\ \emph {et~al.}(1966)\citenamefont {Gobrecht}, \citenamefont {Seeck},\ and\ \citenamefont {Klose}}]{Gobrecht_ZP_190_1966}%
  \BibitemOpen
  \bibfield  {author} {\bibinfo {author} {\bibfnamefont {H.}~\bibnamefont {Gobrecht}}, \bibinfo {author} {\bibfnamefont {S.}~\bibnamefont {Seeck}},\ and\ \bibinfo {author} {\bibfnamefont {T.}~\bibnamefont {Klose}},\ }\bibfield  {title} {\bibinfo {title} {Der {E}influ{\ss} der freien {L}adungstr{\"a}ger auf die optischen {K}onstanten des {Bi$_2$Se$_3$} im {W}ellenl{\"a}ngengebiet von 2 bis 23~$\mu$m},\ }\href {https://doi.org/10.1007/BF01327261} {\bibfield  {journal} {\bibinfo  {journal} {Z. Phys.}\ }\textbf {\bibinfo {volume} {190}},\ \bibinfo {pages} {427} (\bibinfo {year} {1966})}\BibitemShut {NoStop}%
\bibitem [{\citenamefont {Martinez}\ \emph {et~al.}(2017)\citenamefont {Martinez}, \citenamefont {Piot}, \citenamefont {Hakl}, \citenamefont {Potemski}, \citenamefont {Hor}, \citenamefont {Materna}, \citenamefont {Strzelecka}, \citenamefont {Hruban}, \citenamefont {Caha}, \citenamefont {Nov{\'a}k}, \citenamefont {Dubroka}, \citenamefont {Dra{\v{s}}ar},\ and\ \citenamefont {Orlita}}]{Martinez_SR_7_2017}%
  \BibitemOpen
  \bibfield  {author} {\bibinfo {author} {\bibfnamefont {G.}~\bibnamefont {Martinez}}, \bibinfo {author} {\bibfnamefont {B.~A.}\ \bibnamefont {Piot}}, \bibinfo {author} {\bibfnamefont {M.}~\bibnamefont {Hakl}}, \bibinfo {author} {\bibfnamefont {M.}~\bibnamefont {Potemski}}, \bibinfo {author} {\bibfnamefont {Y.~S.}\ \bibnamefont {Hor}}, \bibinfo {author} {\bibfnamefont {A.}~\bibnamefont {Materna}}, \bibinfo {author} {\bibfnamefont {S.~G.}\ \bibnamefont {Strzelecka}}, \bibinfo {author} {\bibfnamefont {A.}~\bibnamefont {Hruban}}, \bibinfo {author} {\bibfnamefont {O.}~\bibnamefont {Caha}}, \bibinfo {author} {\bibfnamefont {J.}~\bibnamefont {Nov{\'a}k}}, \bibinfo {author} {\bibfnamefont {A.}~\bibnamefont {Dubroka}}, \bibinfo {author} {\bibfnamefont {{\v{C}}.}~\bibnamefont {Dra{\v{s}}ar}},\ and\ \bibinfo {author} {\bibfnamefont {M.}~\bibnamefont {Orlita}},\ }\bibfield  {title} {\bibinfo {title} {Determination of the energy band gap of {Bi$_2$Se$_3$}},\ }\href {https://doi.org/10.1038/s41598-017-07211-x} {\bibfield
  {journal} {\bibinfo  {journal} {Sci. Rep.}\ }\textbf {\bibinfo {volume} {7}},\ \bibinfo {pages} {6891} (\bibinfo {year} {2017})}\BibitemShut {NoStop}%
\bibitem [{\citenamefont {Hatch}\ \emph {et~al.}(2011)\citenamefont {Hatch}, \citenamefont {Bianchi}, \citenamefont {Guan}, \citenamefont {Bao}, \citenamefont {Mi}, \citenamefont {Iversen}, \citenamefont {Nilsson}, \citenamefont {Hornek{\ae}r},\ and\ \citenamefont {Hofmann}}]{Hatch_PRB_83_2011}%
  \BibitemOpen
  \bibfield  {author} {\bibinfo {author} {\bibfnamefont {R.~C.}\ \bibnamefont {Hatch}}, \bibinfo {author} {\bibfnamefont {M.}~\bibnamefont {Bianchi}}, \bibinfo {author} {\bibfnamefont {D.}~\bibnamefont {Guan}}, \bibinfo {author} {\bibfnamefont {S.}~\bibnamefont {Bao}}, \bibinfo {author} {\bibfnamefont {J.}~\bibnamefont {Mi}}, \bibinfo {author} {\bibfnamefont {B.~B.}\ \bibnamefont {Iversen}}, \bibinfo {author} {\bibfnamefont {L.}~\bibnamefont {Nilsson}}, \bibinfo {author} {\bibfnamefont {L.}~\bibnamefont {Hornek{\ae}r}},\ and\ \bibinfo {author} {\bibfnamefont {P.}~\bibnamefont {Hofmann}},\ }\bibfield  {title} {\bibinfo {title} {Stability of the {Bi$_2$Se$_3$}(111) topological state: {E}lectron-phonon and electron-defect scattering},\ }\href {https://doi.org/10.1103/PhysRevB.83.241303} {\bibfield  {journal} {\bibinfo  {journal} {Phys. Rev. B}\ }\textbf {\bibinfo {volume} {83}},\ \bibinfo {pages} {241303(R)} (\bibinfo {year} {2011})}\BibitemShut {NoStop}%
\bibitem [{\citenamefont {Pan}\ \emph {et~al.}(2013)\citenamefont {Pan}, \citenamefont {Vescovo}, \citenamefont {Fedorov}, \citenamefont {Gu},\ and\ \citenamefont {Valla}}]{Pan_PRB_88_2013}%
  \BibitemOpen
  \bibfield  {author} {\bibinfo {author} {\bibfnamefont {Z.-H.}\ \bibnamefont {Pan}}, \bibinfo {author} {\bibfnamefont {E.}~\bibnamefont {Vescovo}}, \bibinfo {author} {\bibfnamefont {A.~V.}\ \bibnamefont {Fedorov}}, \bibinfo {author} {\bibfnamefont {G.~D.}\ \bibnamefont {Gu}},\ and\ \bibinfo {author} {\bibfnamefont {T.}~\bibnamefont {Valla}},\ }\bibfield  {title} {\bibinfo {title} {Persistent coherence and spin polarization of topological surface states on topological insulators},\ }\href {https://doi.org/10.1103/PhysRevB.88.041101} {\bibfield  {journal} {\bibinfo  {journal} {Phys. Rev. B}\ }\textbf {\bibinfo {volume} {88}},\ \bibinfo {pages} {041101(R)} (\bibinfo {year} {2013})}\BibitemShut {NoStop}%
\bibitem [{\citenamefont {Nery}\ and\ \citenamefont {Allen}(2016)}]{Nery_PRB_94_2016}%
  \BibitemOpen
  \bibfield  {author} {\bibinfo {author} {\bibfnamefont {J.~P.}\ \bibnamefont {Nery}}\ and\ \bibinfo {author} {\bibfnamefont {P.~B.}\ \bibnamefont {Allen}},\ }\bibfield  {title} {\bibinfo {title} {Influence of {F}r{\"o}hlich polaron coupling on renormalized electron bands in polar semiconductors: {R}esults for zinc-blende {GaN}},\ }\href {https://doi.org/10.1103/PhysRevB.94.115135} {\bibfield  {journal} {\bibinfo  {journal} {Phys. Rev. B}\ }\textbf {\bibinfo {volume} {94}},\ \bibinfo {pages} {115135} (\bibinfo {year} {2016})}\BibitemShut {NoStop}%
\bibitem [{\citenamefont {Chen}\ \emph {et~al.}(2024)\citenamefont {Chen}, \citenamefont {Parker},\ and\ \citenamefont {Monserrat}}]{Chen_PRB_109_2024}%
  \BibitemOpen
  \bibfield  {author} {\bibinfo {author} {\bibfnamefont {S.}~\bibnamefont {Chen}}, \bibinfo {author} {\bibfnamefont {I.~J.}\ \bibnamefont {Parker}},\ and\ \bibinfo {author} {\bibfnamefont {B.}~\bibnamefont {Monserrat}},\ }\bibfield  {title} {\bibinfo {title} {Temperature effects in topological insulators of transition metal dichalcogenide monolayers},\ }\href {https://doi.org/10.1103/PhysRevB.109.155125} {\bibfield  {journal} {\bibinfo  {journal} {Phys. Rev. B}\ }\textbf {\bibinfo {volume} {109}},\ \bibinfo {pages} {155125} (\bibinfo {year} {2024})}\BibitemShut {NoStop}%
\bibitem [{\citenamefont {K{\"o}hler}\ and\ \citenamefont {Becker}(1974)}]{Koehler_PSSB_61_1974}%
  \BibitemOpen
  \bibfield  {author} {\bibinfo {author} {\bibfnamefont {H.}~\bibnamefont {K{\"o}hler}}\ and\ \bibinfo {author} {\bibfnamefont {C.~R.}\ \bibnamefont {Becker}},\ }\bibfield  {title} {\bibinfo {title} {Optically active lattice vibrations in {Bi$_2$Se$_3$}},\ }\href {https://doi.org/10.1002/pssb.2220610218} {\bibfield  {journal} {\bibinfo  {journal} {Phys. Status Solidi B}\ }\textbf {\bibinfo {volume} {61}},\ \bibinfo {pages} {533} (\bibinfo {year} {1974})}\BibitemShut {NoStop}%
\bibitem [{\citenamefont {K{\"{o}}hler}(1973)}]{Koehler_PSSB_58_1973}%
  \BibitemOpen
  \bibfield  {author} {\bibinfo {author} {\bibfnamefont {H.}~\bibnamefont {K{\"{o}}hler}},\ }\bibfield  {title} {\bibinfo {title} {Conduction band parameters of {Bi$_2$Se$_3$} from {S}hubnikov-de {H}aas investigations},\ }\href {https://doi.org/10.1002/pssb.2220580109} {\bibfield  {journal} {\bibinfo  {journal} {Phys. Status Solidi B}\ }\textbf {\bibinfo {volume} {58}},\ \bibinfo {pages} {91} (\bibinfo {year} {1973})}\BibitemShut {NoStop}%
\bibitem [{\citenamefont {Piot}\ \emph {et~al.}(2016)\citenamefont {Piot}, \citenamefont {Desrat}, \citenamefont {Maude}, \citenamefont {Orlita}, \citenamefont {Potemski}, \citenamefont {Martinez},\ and\ \citenamefont {Hor}}]{Piot_PRB_93_2016}%
  \BibitemOpen
  \bibfield  {author} {\bibinfo {author} {\bibfnamefont {B.~A.}\ \bibnamefont {Piot}}, \bibinfo {author} {\bibfnamefont {W.}~\bibnamefont {Desrat}}, \bibinfo {author} {\bibfnamefont {D.~K.}\ \bibnamefont {Maude}}, \bibinfo {author} {\bibfnamefont {M.}~\bibnamefont {Orlita}}, \bibinfo {author} {\bibfnamefont {M.}~\bibnamefont {Potemski}}, \bibinfo {author} {\bibfnamefont {G.}~\bibnamefont {Martinez}},\ and\ \bibinfo {author} {\bibfnamefont {Y.~S.}\ \bibnamefont {Hor}},\ }\bibfield  {title} {\bibinfo {title} {Hole {F}ermi surface in {Bi$_2$Se$_3$} probed by quantum oscillations},\ }\href {https://doi.org/10.1103/physrevb.93.155206} {\bibfield  {journal} {\bibinfo  {journal} {Phys. Rev. B}\ }\textbf {\bibinfo {volume} {93}},\ \bibinfo {pages} {155206} (\bibinfo {year} {2016})}\BibitemShut {NoStop}%
\bibitem [{\citenamefont {Trushin}\ and\ \citenamefont {G{\"o}rling}(2018)}]{Trushin_PRL_120_2018}%
  \BibitemOpen
  \bibfield  {author} {\bibinfo {author} {\bibfnamefont {E.}~\bibnamefont {Trushin}}\ and\ \bibinfo {author} {\bibfnamefont {A.}~\bibnamefont {G{\"o}rling}},\ }\bibfield  {title} {\bibinfo {title} {Topological phase transitions in zinc-blende semimetals driven exclusively by electronic temperature},\ }\href {https://doi.org/10.1103/PhysRevLett.120.146401} {\bibfield  {journal} {\bibinfo  {journal} {Phys. Rev. Lett.}\ }\textbf {\bibinfo {volume} {120}},\ \bibinfo {pages} {146401} (\bibinfo {year} {2018})}\BibitemShut {NoStop}%
\bibitem [{\citenamefont {Karasiev}\ \emph {et~al.}(2018)\citenamefont {Karasiev}, \citenamefont {Dufty},\ and\ \citenamefont {Trickey}}]{Karasiev_PRL_120_2018}%
  \BibitemOpen
  \bibfield  {author} {\bibinfo {author} {\bibfnamefont {V.~V.}\ \bibnamefont {Karasiev}}, \bibinfo {author} {\bibfnamefont {J.~W.}\ \bibnamefont {Dufty}},\ and\ \bibinfo {author} {\bibfnamefont {S.~B.}\ \bibnamefont {Trickey}},\ }\bibfield  {title} {\bibinfo {title} {Nonempirical semilocal free-energy density functional for matter under extreme conditions},\ }\href {https://doi.org/10.1103/PhysRevLett.120.076401} {\bibfield  {journal} {\bibinfo  {journal} {Phys. Rev. Lett.}\ }\textbf {\bibinfo {volume} {120}},\ \bibinfo {pages} {076401} (\bibinfo {year} {2018})}\BibitemShut {NoStop}%
\bibitem [{\citenamefont {Verstraete}\ \emph {et~al.}(2025)\citenamefont {Verstraete}, \citenamefont {Abreu}, \citenamefont {Allemand}, \citenamefont {Amadon}, \citenamefont {Antonius}, \citenamefont {Azizi}, \citenamefont {Baguet}, \citenamefont {Barat}, \citenamefont {Bastogne}, \citenamefont {B{\'e}jaud}, \citenamefont {Beuken}, \citenamefont {Bieder}, \citenamefont {Blanchet}, \citenamefont {Bottin}, \citenamefont {Bouchet}, \citenamefont {Bouquiaux}, \citenamefont {Bousquet}, \citenamefont {Boust}, \citenamefont {Brieuc}, \citenamefont {Brousseau-Couture}, \citenamefont {Brouwer}, \citenamefont {Bruneval}, \citenamefont {Castellano}, \citenamefont {Castiel}, \citenamefont {Charraud}, \citenamefont {Cl{\'e}rouin}, \citenamefont {C{\^o}t{\'e}}, \citenamefont {Duval}, \citenamefont {Gallo}, \citenamefont {Gendron}, \citenamefont {Geneste}, \citenamefont {Ghosez}, \citenamefont {Giantomassi}, \citenamefont {Gingras}, \citenamefont {G{\'o}mez-Ortiz}, \citenamefont {Gonze}, \citenamefont {Goudreault},
  \citenamefont {Gr{\"u}neis}, \citenamefont {Gupta}, \citenamefont {Guster}, \citenamefont {Hamann}, \citenamefont {He}, \citenamefont {Hellman}, \citenamefont {Holzwarth}, \citenamefont {Jollet}, \citenamefont {Kestener}, \citenamefont {Lygatsika}, \citenamefont {Nadeau}, \citenamefont {MacEnulty}, \citenamefont {Marazzi}, \citenamefont {Mignolet}, \citenamefont {O'Regan}, \citenamefont {Outerovitch}, \citenamefont {Paillard}, \citenamefont {Petretto}, \citenamefont {Ponc{\'e}}, \citenamefont {Ricci}, \citenamefont {Rignanese}, \citenamefont {Rodriguez-Mayorga}, \citenamefont {Romero}, \citenamefont {Rostami}, \citenamefont {Royo}, \citenamefont {Sarraute}, \citenamefont {Sasani}, \citenamefont {Soubiran}, \citenamefont {Stengel}, \citenamefont {Tantardini}, \citenamefont {Torrent}, \citenamefont {Trinquet}, \citenamefont {Vasilchenko}, \citenamefont {Waroquiers}, \citenamefont {Zabalo}, \citenamefont {Zadoks}, \citenamefont {Zhang},\ and\ \citenamefont {Zwanziger}}]{Verstraete_JCP_163_2025}%
  \BibitemOpen
  \bibfield  {author} {\bibinfo {author} {\bibfnamefont {M.~J.}\ \bibnamefont {Verstraete}}, \bibinfo {author} {\bibfnamefont {J.}~\bibnamefont {Abreu}}, \bibinfo {author} {\bibfnamefont {G.~E.}\ \bibnamefont {Allemand}}, \bibinfo {author} {\bibfnamefont {B.}~\bibnamefont {Amadon}}, \bibinfo {author} {\bibfnamefont {G.}~\bibnamefont {Antonius}}, \bibinfo {author} {\bibfnamefont {M.}~\bibnamefont {Azizi}}, \bibinfo {author} {\bibfnamefont {L.}~\bibnamefont {Baguet}}, \bibinfo {author} {\bibfnamefont {C.}~\bibnamefont {Barat}}, \bibinfo {author} {\bibfnamefont {L.}~\bibnamefont {Bastogne}}, \bibinfo {author} {\bibfnamefont {R.}~\bibnamefont {B{\'e}jaud}}, \bibinfo {author} {\bibfnamefont {J.-M.}\ \bibnamefont {Beuken}}, \bibinfo {author} {\bibfnamefont {J.}~\bibnamefont {Bieder}}, \bibinfo {author} {\bibfnamefont {A.}~\bibnamefont {Blanchet}}, \bibinfo {author} {\bibfnamefont {F.}~\bibnamefont {Bottin}}, \bibinfo {author} {\bibfnamefont {J.}~\bibnamefont {Bouchet}}, \bibinfo {author} {\bibfnamefont
  {J.}~\bibnamefont {Bouquiaux}}, \bibinfo {author} {\bibfnamefont {E.}~\bibnamefont {Bousquet}}, \bibinfo {author} {\bibfnamefont {J.}~\bibnamefont {Boust}}, \bibinfo {author} {\bibfnamefont {F.}~\bibnamefont {Brieuc}}, \bibinfo {author} {\bibfnamefont {V.}~\bibnamefont {Brousseau-Couture}}, \bibinfo {author} {\bibfnamefont {N.}~\bibnamefont {Brouwer}}, \bibinfo {author} {\bibfnamefont {F.}~\bibnamefont {Bruneval}}, \bibinfo {author} {\bibfnamefont {A.}~\bibnamefont {Castellano}}, \bibinfo {author} {\bibfnamefont {E.}~\bibnamefont {Castiel}}, \bibinfo {author} {\bibfnamefont {J.-B.}\ \bibnamefont {Charraud}}, \bibinfo {author} {\bibfnamefont {J.}~\bibnamefont {Cl{\'e}rouin}}, \bibinfo {author} {\bibfnamefont {M.}~\bibnamefont {C{\^o}t{\'e}}}, \bibinfo {author} {\bibfnamefont {C.}~\bibnamefont {Duval}}, \bibinfo {author} {\bibfnamefont {A.}~\bibnamefont {Gallo}}, \bibinfo {author} {\bibfnamefont {F.}~\bibnamefont {Gendron}}, \bibinfo {author} {\bibfnamefont {G.}~\bibnamefont {Geneste}}, \bibinfo {author}
  {\bibfnamefont {P.}~\bibnamefont {Ghosez}}, \bibinfo {author} {\bibfnamefont {M.}~\bibnamefont {Giantomassi}}, \bibinfo {author} {\bibfnamefont {O.}~\bibnamefont {Gingras}}, \bibinfo {author} {\bibfnamefont {F.}~\bibnamefont {G{\'o}mez-Ortiz}}, \bibinfo {author} {\bibfnamefont {X.}~\bibnamefont {Gonze}}, \bibinfo {author} {\bibfnamefont {F.~A.}\ \bibnamefont {Goudreault}}, \bibinfo {author} {\bibfnamefont {A.}~\bibnamefont {Gr{\"u}neis}}, \bibinfo {author} {\bibfnamefont {R.}~\bibnamefont {Gupta}}, \bibinfo {author} {\bibfnamefont {B.}~\bibnamefont {Guster}}, \bibinfo {author} {\bibfnamefont {D.~R.}\ \bibnamefont {Hamann}}, \bibinfo {author} {\bibfnamefont {X.}~\bibnamefont {He}}, \bibinfo {author} {\bibfnamefont {O.}~\bibnamefont {Hellman}}, \bibinfo {author} {\bibfnamefont {N.}~\bibnamefont {Holzwarth}}, \bibinfo {author} {\bibfnamefont {F.}~\bibnamefont {Jollet}}, \bibinfo {author} {\bibfnamefont {P.}~\bibnamefont {Kestener}}, \bibinfo {author} {\bibfnamefont {I.-M.}\ \bibnamefont {Lygatsika}}, \bibinfo
  {author} {\bibfnamefont {O.}~\bibnamefont {Nadeau}}, \bibinfo {author} {\bibfnamefont {L.}~\bibnamefont {MacEnulty}}, \bibinfo {author} {\bibfnamefont {E.}~\bibnamefont {Marazzi}}, \bibinfo {author} {\bibfnamefont {M.}~\bibnamefont {Mignolet}}, \bibinfo {author} {\bibfnamefont {D.~D.}\ \bibnamefont {O'Regan}}, \bibinfo {author} {\bibfnamefont {R.}~\bibnamefont {Outerovitch}}, \bibinfo {author} {\bibfnamefont {C.}~\bibnamefont {Paillard}}, \bibinfo {author} {\bibfnamefont {G.}~\bibnamefont {Petretto}}, \bibinfo {author} {\bibfnamefont {S.}~\bibnamefont {Ponc{\'e}}}, \bibinfo {author} {\bibfnamefont {F.}~\bibnamefont {Ricci}}, \bibinfo {author} {\bibfnamefont {G.-M.}\ \bibnamefont {Rignanese}}, \bibinfo {author} {\bibfnamefont {M.}~\bibnamefont {Rodriguez-Mayorga}}, \bibinfo {author} {\bibfnamefont {A.~H.}\ \bibnamefont {Romero}}, \bibinfo {author} {\bibfnamefont {S.}~\bibnamefont {Rostami}}, \bibinfo {author} {\bibfnamefont {M.}~\bibnamefont {Royo}}, \bibinfo {author} {\bibfnamefont {M.}~\bibnamefont
  {Sarraute}}, \bibinfo {author} {\bibfnamefont {A.}~\bibnamefont {Sasani}}, \bibinfo {author} {\bibfnamefont {F.}~\bibnamefont {Soubiran}}, \bibinfo {author} {\bibfnamefont {M.}~\bibnamefont {Stengel}}, \bibinfo {author} {\bibfnamefont {C.}~\bibnamefont {Tantardini}}, \bibinfo {author} {\bibfnamefont {M.}~\bibnamefont {Torrent}}, \bibinfo {author} {\bibfnamefont {V.}~\bibnamefont {Trinquet}}, \bibinfo {author} {\bibfnamefont {V.}~\bibnamefont {Vasilchenko}}, \bibinfo {author} {\bibfnamefont {D.}~\bibnamefont {Waroquiers}}, \bibinfo {author} {\bibfnamefont {A.}~\bibnamefont {Zabalo}}, \bibinfo {author} {\bibfnamefont {A.}~\bibnamefont {Zadoks}}, \bibinfo {author} {\bibfnamefont {H.}~\bibnamefont {Zhang}},\ and\ \bibinfo {author} {\bibfnamefont {J.}~\bibnamefont {Zwanziger}},\ }\bibfield  {title} {\bibinfo {title} {{Abinit} 2025: {N}ew capabilities for the predictive modeling of solids and nanomaterials},\ }\href {https://doi.org/10.1063/5.0288278} {\bibfield  {journal} {\bibinfo  {journal} {J. Chem. Phys.}\
  }\textbf {\bibinfo {volume} {163}},\ \bibinfo {pages} {164126} (\bibinfo {year} {2025})}\BibitemShut {NoStop}%
\bibitem [{\citenamefont {Torrent}\ \emph {et~al.}(2008)\citenamefont {Torrent}, \citenamefont {Jollet}, \citenamefont {Bottin}, \citenamefont {Z{\'e}rah},\ and\ \citenamefont {Gonze}}]{Torrent_CMS_42_2008}%
  \BibitemOpen
  \bibfield  {author} {\bibinfo {author} {\bibfnamefont {M.}~\bibnamefont {Torrent}}, \bibinfo {author} {\bibfnamefont {F.}~\bibnamefont {Jollet}}, \bibinfo {author} {\bibfnamefont {F.}~\bibnamefont {Bottin}}, \bibinfo {author} {\bibfnamefont {G.}~\bibnamefont {Z{\'e}rah}},\ and\ \bibinfo {author} {\bibfnamefont {X.}~\bibnamefont {Gonze}},\ }\bibfield  {title} {\bibinfo {title} {Implementation of the projector augmented-wave method in the {ABINIT} code: {A}pplication to the study of iron under pressure},\ }\href {https://doi.org/10.1016/j.commatsci.2007.07.020} {\bibfield  {journal} {\bibinfo  {journal} {Comput. Mater. Sci.}\ }\textbf {\bibinfo {volume} {42}},\ \bibinfo {pages} {337} (\bibinfo {year} {2008})}\BibitemShut {NoStop}%
\bibitem [{\citenamefont {Rubel}\ and\ \citenamefont {Sadikian}(2026)}]{Rubel_DFT2026}%
  \BibitemOpen
  \bibfield  {author} {\bibinfo {author} {\bibfnamefont {O.}~\bibnamefont {Rubel}}\ and\ \bibinfo {author} {\bibfnamefont {K.}~\bibnamefont {Sadikian}},\ }\bibfield  {title} {\bibinfo {title} {Topological robustness of thermally disordered lattices: {I}nstantaneous symmetry breaking and band gap fluctuations},\ }in\ \href@noop {} {\emph {\bibinfo {booktitle} {21st International Conference on Density Functional Theory and its Applications}}}\ (\bibinfo {year} {2026})\ \bibinfo {note} {(submitted May 7, 2026; publicly available Jul 31, 2026)}\BibitemShut {NoStop}%
\bibitem [{\citenamefont {Lu}\ and\ \citenamefont {Sun}(2026)}]{Lu_N_x_2026}%
  \BibitemOpen
  \bibfield  {author} {\bibinfo {author} {\bibfnamefont {L.}~\bibnamefont {Lu}}\ and\ \bibinfo {author} {\bibfnamefont {Y.-Y.}\ \bibnamefont {Sun}},\ }\bibfield  {title} {\bibinfo {title} {Weakened spin-orbit coupling drives topological-to-normal-insulator transition at elevated temperatures},\ }\href {https://doi.org/10.1016/j.newton.2026.100621} {\bibfield  {journal} {\bibinfo  {journal} {Newton}\ }\textbf {\bibinfo {volume} {2}},\ \bibinfo {pages} {100621} (\bibinfo {year} {2026})},\ \bibinfo {note} {{DOI}: 10.1016/j.newton.2026.100621}\BibitemShut {NoStop}%
\end{thebibliography}%


\end{document}